\documentclass{article}
\usepackage[a4paper, total={6.5in, 10in}]{geometry}
\usepackage{blindtext}

\usepackage{amsmath}
\usepackage{enumerate}

\usepackage{xcolor}
\usepackage{colortbl}
\usepackage{tabularray}
\usepackage{tikz}

\usepackage{graphicx}
\usepackage{siunitx}
\usepackage{hyperref}
\usepackage{subcaption}
\usepackage[affil-it]{authblk}

\newcommand*\circled[1]{\tikz[baseline=(char.base)]{
            \node[shape=circle,draw,inner sep=1pt] (char) {#1};}}  

\graphicspath{{./figure/}{./figure/rod-results/}{./figure/bracket-results/}{./figure/bike-results/}{./figure/piston-results/}}

\title{Model Consistency for Mechanical Design: \\ Bridging Lumped and Distributed Parameter Models with A Priori Guarantees}
\author[1]{Randi Wang\thanks{corresponding author: rwang@parc.com, moradbeh@parc.com}}
\author[2]{Vadim Shapiro}
\author[3]{Morad Behandish$^*$}

\affil[1]{%
    Palo Alto Research Center (PARC), Palo Alto, CA }
\affil[2]{%
    International Computer Science Institute (ICSI), Berkeley, CA}
\affil[3]{%
    Palo Alto Research Center (PARC), Palo Alto, CA }

\date{\vspace{-5ex}}

\begin{document}
\maketitle

\begin{abstract}
Engineering design often involves representation in at least two levels of abstraction: the system-level, represented by lumped parameter models (LPMs), and the geometric-level, represented by distributed parameter models (DPMs). Functional design innovation commonly occurs at the system-level, followed by a geometric-level realization of functional LPM components. However, comparing these two levels in terms of behavioral outcomes can be challenging and time-consuming, leading to delays in design translations between system and mechanical engineers.
In this paper, we propose a simulation-free scheme that compares LPMs and spatially-discretized DPMs based on their model specifications and behaviors of interest, regardless of modeling languages and numerical methods. We adopt a model order reduction (MOR) technique that a priori guarantees accuracy, stability, and convergence to improve the computational efficiency of large-scale models. Our approach is demonstrated through the model consistency analysis of several mechanical designs, showing its validity, efficiency, and generality. Our method provides a systematic way to compare system-level and geometric-level designs, improving reliability and facilitating design translation.
\end{abstract}


\section{Introduction}

In the field of engineering design, the initial step towards creating innovative designs often involves system design, which entails conceptualizing and developing functional structures. Once a system is designed, the subsequent step involves developing its 3D geometry for manufacturing purposes. The objective of the system-based geometric design process is to identify at least one geometric realization that matches the system design. In this paper, we restrict our attention to mechanical systems whose representation at the system-level can be given by mass-spring-damper networks.

\subsection{Motivation}
The DPMs (e.g., 3D geometric assemblies) and LPMs (e.g., mechanical mass-spring-damper networks) that describe the same physical system at different levels of abstraction are represented using different languages and semantics that cannot be directly translated into one another. For instance, Dassault Systèmes offers two commercial software, Dymola for system modeling and SolidWorks for geometric modeling. The models created in these programs are incompatible and cannot be automatically translated into each other owing to disparities in model types and representations \cite{systemes2018dymola, chen2022graph,almattar2019learn,chen2020maximal}. This gap presents a significant challenge for ensuring consistency between the system models and computer-aided design/engineering (CAD/CAE) models. This paper concentrates on the crucial technical issue of systematically verifying the consistency between the two models.

LPMs use a network of lumped components to represent the structure of engineering systems, such as masses, springs, and dampers, which are governed by a system of ordinary differential equations (ODEs) in terms of variables that vary with time \cite{karnopp2012system,wang2019topological}. On the other hand, DPMs explicitly consider the geometric and material properties of engineering systems, which are governed by a system of partial differential equations (PDEs) that take into account variables varying with both time and space \cite{mazumder2015numerical}. Numerical discretization methods can be used to approximate these PDEs by large systems of ODEs upon approximating the spatial continuum in a finite basis \cite{mattiussi2000finite}.

\begin{figure}[!htb]
        \centering
        \includegraphics[width=0.8\linewidth]{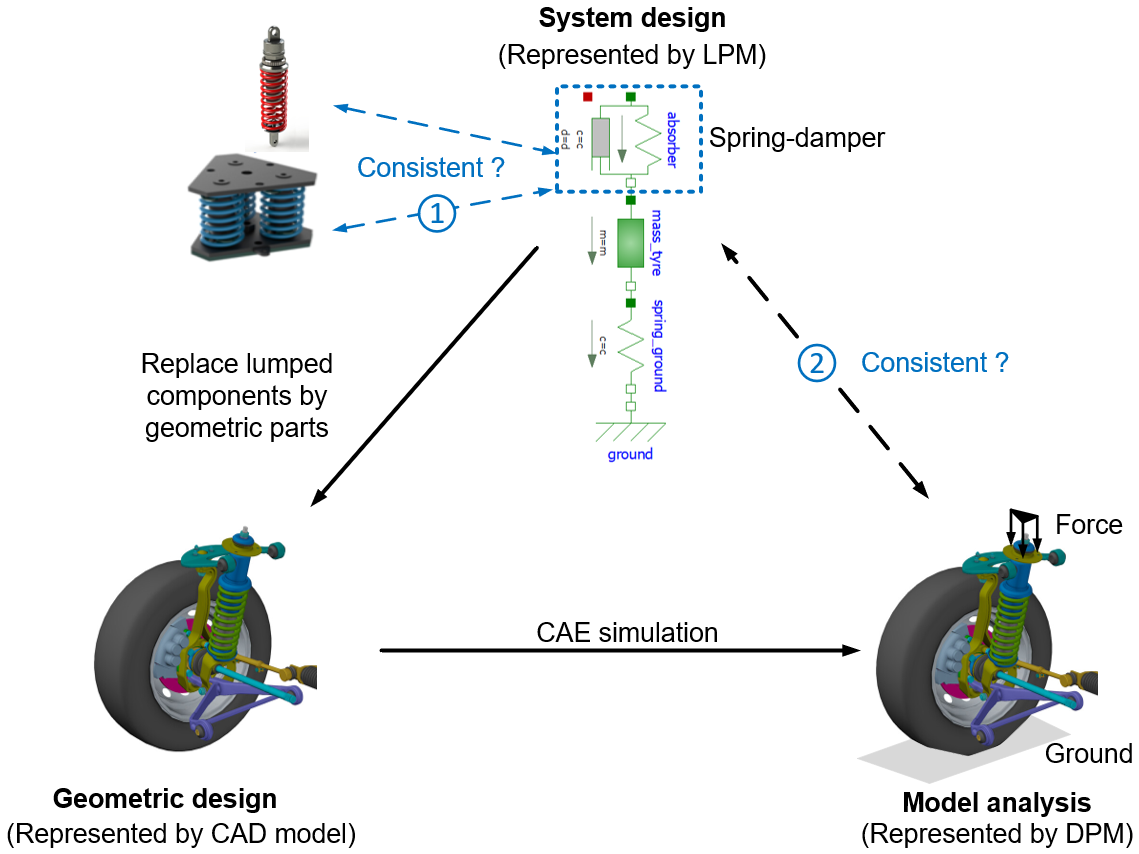}
        \caption{The role of model consistency analysis in the process of system-based geometric design}
        \label{fig:motivation_consistency}
    \end{figure}

Figure \ref{fig:motivation_consistency} depicts the process of designing a 3D suspension mechanism based on a system design. After creating an LPM in Modelica \cite{fritzson2011introduction}, mechanical parts are used to realize the lumped components and obtain the expected behaviors.  The stiffness and damping of the absorber are modeled by a spring-damper pair, without considering the precise geometric realization.  Two potential geometric realization options for the absorber are presented in the library. While there might be various options for the absorber, a designer must verify that the selected option behaves as intended by the lumped component (\circled{1} in the figure) before the assembly process or design qualified parts guided by the three model consistency conditions explained in Section \ref{sec:formulation}. Additionally, it is crucial to confirm that the final geometric assembly behaves as intended by the LPM (depicted as \circled{2} in the figure) to ensure a qualified design.   Currently, the only reliable way to compare design behaviors is to simulate both the LPM and DPM and compare their differential equation solutions a posteriori, which is computationally prohibitive for large-scale models, not to mention the additional challenges related to selecting appropriate time-steps, stability, and convergence. The goal of this paper is to propose a systematic method to check consistency between the system models and CAD/CAE models.

\subsection{Contributions}

Our main technical contributions are:

\begin{enumerate} 
    \item We propose a generalizable definition of {\it consistency} between mechanical LPMs and DPMs that considers both model {\it specifications} and {\it solutions}, taking into account factors such as mass, initial and boundary conditions (ICs/BCs), and the behavior of interest (BoI). 
    \item We develop a {\it simulation-free} scheme for checking the consistency between LPMs and DPMs based on the definition proposed above. This idea is to compute a priori error bounds between the solutions of both model types by comparing their parameters, circumventing the costly process of solving differential equations.    
    \item We adopt a MOR technique that a priori guarantees accuracy, stability, and convergence in the simulation-free scheme to improve the efficiency of consistency analysis for large-scale models. The MOR enables the analysis of large-scale designs in a computationally efficient manner. 
\end{enumerate}

\subsection{Outline} 
In Section \ref{sec:related_works}, we review the prior works on system-based geometric design, physical model solution comparison, and MOR. In Section \ref{sec:formulation}, we define the problem of model consistency analysis. In Section \ref{sec:scheme}, we present a simulation-free scheme to compare LPMs and DPMs. We illustrate the application of our scheme to various mechanical problems in Section \ref{sec:application}. Finally, in Section \ref{sec:conclusion}, we summarize the implications of our proposed scheme and suggest potential future research directions.

\section{Related Work} \label{sec:related_works}

\subsection{System-Based Geometric Design}\label{section:reivewdesign}

The system-to-geometry design strategy has been shown by Ulrich \cite {ulrich2003product} to reduce complexity and clarify the problem-solving process. Most system-based geometric design approaches rely on a one-to-one correspondence between the lumped component in the LPM and the geometric part in a solid model repository. For instance, Finger and Rinderle proposed a component database that contains correspondence between ports of bond graphs and geometric parts for a specific class of mechanical designs \cite{finger1990transformational}. 
 Engelson developed an integrated environment that combines geometric design and system modeling tools to assist engineers in constructing and verifying large, moving rigid-body assemblies \cite{engelson2003mechanical}.  However, these approaches may result in unrealistic designs by ignoring {\it function-sharing}. Ulrich's work stands out as an exception \cite{ulrich2003product}, as it demonstrates a systematic approach to merge multiple functions, abstracted by different system-level components, in fewer geometric parts. Despite the numerous approaches proposed for system-based geometric design \cite{prabhu1989synthesis,greer2002effort,ulrich2003product,engelson2003mechanical,kota1995designing,greer2004effort}, none of them formally introduce the concept of consistency between system and geometric designs, nor do they provide a systematic approach for assessing the validity of the geometric design with respect to the target system behavior.

\subsection{Solution Comparison between LPM and DPM}\label{section:reivewbehaviors}

Comparing the solutions of LPMs and DPMs is a common requirement in model conversion problems, where the two models must be converted to each other with tolerable differences in their respective solutions.
In computer graphics, for instance, deformable objects like cloth fabrics and soft tissues are often converted to lumped mass-spring models for faster simulations due to the simplicity and efficiency of LPMs \cite{mollemans2004fast,baraff1998large,kahler2001geometry}. Gelder \cite{gelder1998approximate} developed a lumped-spring element based on the geometric angle and length information of 2D linear triangular finite elements, while Vincent et al. \cite{baudet2007new} extended the method to rectangular finite elements. In both cases, the solutions of the converted LPM and the original finite element model can be directly compared.  Suriya  et al. in \cite{natsupakpong2010determination} proposed a method for converting 3D deformable objects from finite element models to LPMs by minimizing the difference of the stiffness matrices of the two models. However, the dimension of the LPM solution is often much smaller than that of the DPM after spatial discretization, making direct comparison challenging. In other words, a one-to-one correspondence between the solution dimensions of the two models is typically not guaranteed. This limits the applicability of existing methods to the model solution comparison problem addressed in this paper.

\subsection{Model Order Reduction}
The field of MOR has seen the development of various techniques \cite{baur2014model, fang2022110823, qu2004model, huang2022applications, fang2023113897} to approximate a given ``full-order'' model (FOM) in a numerically efficient and stable manner while preserving certain desired properties. The balanced truncation (BT) method, for instance, removes weakly controllable and observable states from the FOM \cite{chahlaoui2005model}, but can be computationally expensive for large-scale models \cite{besselink2013comparison}. The rational Krylov subspace (RKS) method, on the other hand, approximates the FOM transfer function by matching a few significant terms of its Taylor series expansion \cite{lohmann2000introduction}, but does not have a standard rule for choosing the frequency, which is the frequency around which the Taylor series expansion is made. To address this limitation, the iterative rational Krylov algorithm (IRKA) was developed, which iteratively updates the expansion frequency using the reflection of poles of the updated ROM about the imaginary axis at each iteration step until the difference between the transfer functions of FOM and ROM is minimized \cite{gugercin2008h_2}. 

However, the IRKA algorithm cannot ensure the error converges to a local minimum \cite{beattie2009trust}. To overcome this limitation, the CUmulative REduction (CURE) scheme was proposed, which adaptively chooses the expansion frequency and incrementally increases the scale of the ROM by monotonically decreasing the norm of the error transfer function to zero through an accumulation process \cite{panzer2014model}. Furthermore, a stability-preserving, adaptive rational Krylov (SPARK) algorithm \cite{panzer2014model} was developed to maintain model stability and is usually embedded in the CURE scheme to generate a family of stable ROMs whose orders are increased by sequential accumulation in a single MOR process. This embedded CURE scheme with SPARK algorithm (hereafter abbreviated by SPARK+CURE) was used in \cite{wang2022surrogate} to generate a family of physically-interpretable multi-fidelity surrogate LPMs for physical systems governed by PDEs.

The SPARK+CURE method has several key advantages; namely: 1) automatic search for proper expansion frequencies; 2) preserved model stability; 3) guaranteed error convergence; and 4) a priori error bound. This method is particularly suitable for large-scale mechanical systems, as it can significantly improve the time efficiency of model consistency analysis. In Table \ref{fig:MOR-comparison}, we compare the important properties of the BT, RKS, IRKA, and SPARK+CURE methods, highlighting the advantages of the latter. For our proposed simulation-free scheme (Section \ref{sec:scheme}), we adopt the SPARK+CURE method to ensure accuracy, stability, and convergence in our model consistency analysis.

\begin{table}[!htb]
\centering
\caption{Comparison of four popular MOR methods}
\begin{tabular}{|c|c|c|c|>{\columncolor[RGB]{230, 242, 255}}c|}
\hline
\textbf{} & \textbf{BT}  & \textbf{RKS}  & \textbf{IRKA} & \textbf{SPARK+CURE} \\ \hline
\textbf{\begin{tabular}[c]{@{}c@{}}Numerical \\ Efficiency\end{tabular}}  & LOW & HIGH & HIGH & HIGH       \\ \hline
\textbf{\begin{tabular}[c]{@{}c@{}}A priori \\ error bound\end{tabular}}        & YES & NO   & NO   & YES        \\ \hline
\textbf{\begin{tabular}[c]{@{}c@{}}Auto order \\ decision\end{tabular}}   & NO  & NO   & NO   & YES        \\ \hline
\textbf{\begin{tabular}[c]{@{}c@{}}Maintain \\ stability\end{tabular}}    & YES & NO   & YES  & YES        \\ \hline
\textbf{\begin{tabular}[c]{@{}c@{}}Guarantee \\ convergence\end{tabular}} & YES & NO   & NO   & YES        \\ \hline
\end{tabular}\label{fig:MOR-comparison}
\end{table}

\section{Problem formulation} \label{sec:formulation}
This section outlines the problem formulation for the model consistency analysis procedure, which involves two model types: the LPM, which provides a system-level description of physical systems, and the DPM, which incorporates the spatiotemporal distribution of materials. The definition of these models and the relationship between their model specifications, behaviors, and ICs/BCs are crucial for formulating the problem, and will be discussed below.

\subsection{Definitions}
We focus on linear time-invariant (LTI) translational mechanical LPMs that can be modeled as networks of interconnected lumped components such as masses, springs, and dampers. To uniquely determine the system behavior, initial mass displacements, velocities, and source terms must be specified, which give rise to a system of ODEs in the time domain. We assume that any physically meaningful mass-spring-damper network can be realized by at least one DPM, which has a continuous material distribution and geometry embedded in 3D space. The DPM can be divided into pieces such that there is a one-to-one correspondence between the spatial integration of mass density and lumped masses. The effective stiffness and damping of the DPM correspond to those of the mass-spring-damper network, and the spatial integration of ICs/BCs and body effects of different pieces correspond to the ICs and source terms of lumped masses. The DPM behavior is described by PDEs defined over a region of spacetime. To compute the effective stiffness and damping, solving PDEs is usually unavoidable. The solution generally involves spatial discretization and numerical methods such as finite difference, element, and volume analysis \cite{mattiussi2000finite}.

Based on the above definitions, we can define {\it consistency} between an LPM and a DPM as the satisfaction of the following three conditions:

\begin{enumerate}[(C1)]

\item The total lumped mass value of the LPM matches the spatial integration of the mass density of the DPM.

\item The spatial integration of the ICs/BCs, as well as the body effects of the DPM, match the ICs and source terms of the LPM.

\item The BoI of the DPM matches the BoI of the LPM in terms of spatial integration.

\end{enumerate}

In other words, for an LPM and a DPM to be consistent, the mass, ICs, source terms, and BoI of the LPM must be equivalent to the mass distribution, initial and boundary conditions, and BoI of the DPM, respectively.

\subsection{A mechanical example} \label{sec:a_me_ex}

\begin{figure}[!htb]
        \centering
        \includegraphics[width=0.8\linewidth]{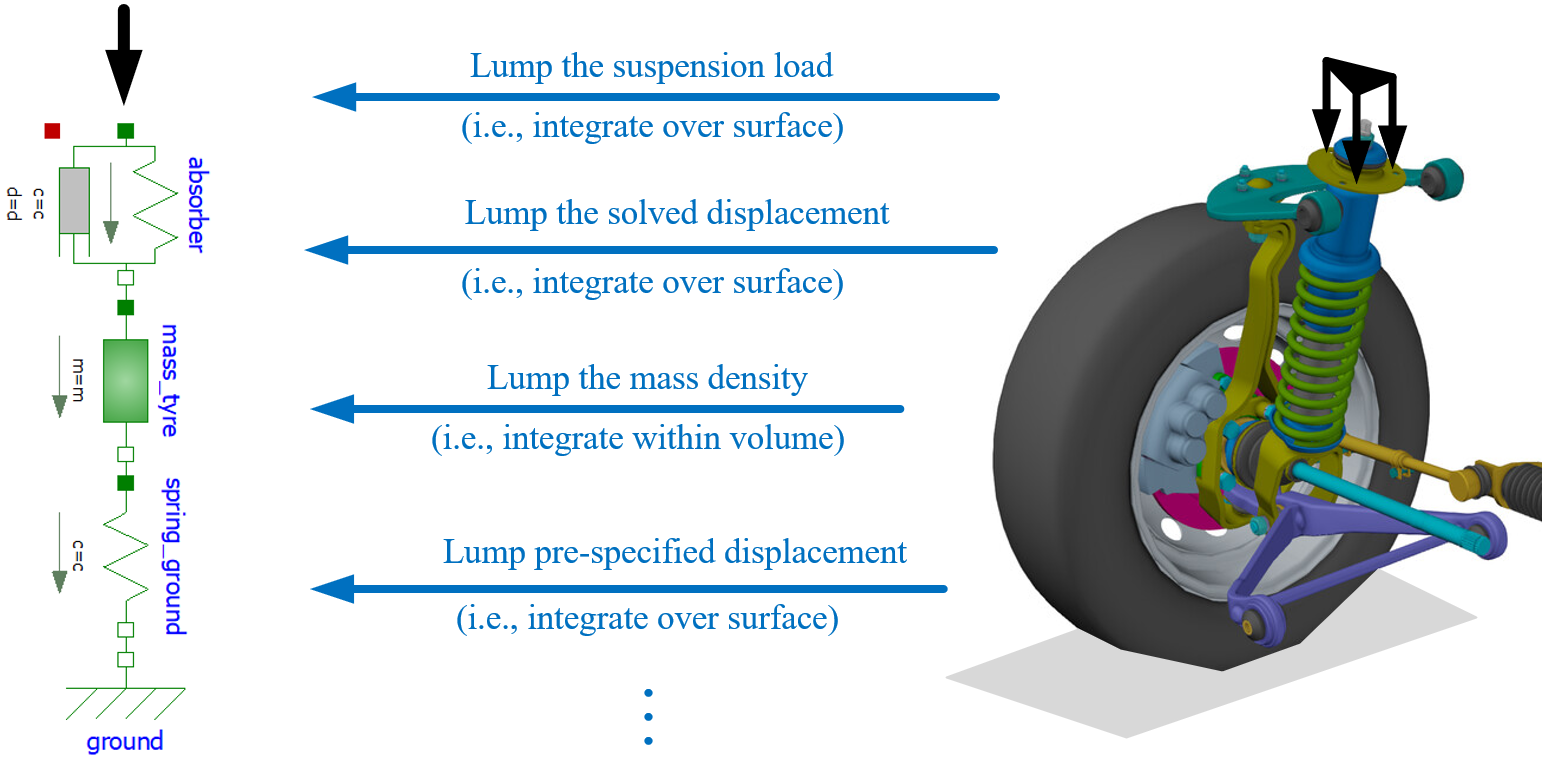}
        \caption{An LPM, its DPM realization, and the model specification match}
        \label{fig:Ts_tyre}
    \end{figure}

Figure \ref{fig:Ts_tyre} depicts an example of an LPM and its corresponding DPM realization, where the geometry of a 3D suspension system is considered. The LPM is composed of a mass, two springs, and a damper, and is subjected to a time-varying external force ${\bf{f}}(t)$. The network is connected to the ground via a spring. On the other hand, the DPM is an assembly of several solid parts that have linear-elastic material distributed in 3D space. The top surface of the assembly is subjected to time-varying pressure ${\bf{p}}({\bf{x}},t)$, where $\mathbf{x} $ is the spatial 3D coordinates. The entire assembly is fixed to the ground. To ensure consistency between the LPM and DPM, we require that the following conditions are satisfied: the mass of the DPM matches the total mass of the LPM,  i.e., $m=\int {\rho({\bf{x}})}dV$  where $\rho({\bf{x}})$  is the material density and $dV$ represents an infinitesimal volume element; the external load on the LPM matches the total force on the DPM, i.e., ${\bf{f}}(t)=\int {\bf{p}}({\bf{x}},t)dS$ where $dS$ represents an infinitesimal surface element; the pre-specified displacement $\bar{\bf{u}}({\bf{x}},t)$ of the LPM (which is fixed to the ground) matches that of the DPM, i.e., $\bar{\bf{w}}(t)=\frac{1}{S_c}\int \bar{\bf{u}}({\bf{x}},t)dS={\bf{0}}$, where $S_c$ is the contact surface area between the tire and the ground; and the solved displacement of the LPM matches that of the DPM, i.e., ${\bf{w}}(t)=\frac{1}{S_p}\int {\bf{u}}({\bf{x}},t)dS$, where $S_p$ is the top surface area of the DPM assembly.

\subsection{Formulation}
To analyze consistency between LPM and DPM, one has to verify if the conditions (C1) through (C3) are satisfied. 

For practical purposes, we introduce an upper-bound (i.e., inequality constraint) to relax the strict equality constraint of BoI in (C3). Noting that an LPM is a ROM of a DPM, its solution can be at best an approximation of the DPM solution, and hence, it is not reasonable to expect a strict equivalence between the solutions of LPM and DPM. 

Given an LPM and a DPM with matched mass properties, ICs/BCs, and source terms, our objective is to check if the overall error between pairs of LPM BoI quantities ${\bf{w}}_i (t)$ and the corresponding DPM BoI quantities ${\bf{u}}_i (t)$ for $i = 1, 2, \ldots, n$ is less than an acceptable tolerance $\epsilon > 0$. Symbolically, this can be represented as $\sum_{i=1}^{n}\left \| {\bf{u}}_i (t) - {\bf{w}}_i (t) \right \|\leq \epsilon $. Any commonly used vector norm such as $L^1$, $L^2$, and $L^{\infty}$ can be used as the norm $\left \| \cdot  \right \|$. However, for the sake of specificity, we use $L^{\infty}$ to provide a measure of the maximum deviation between the DPM and LPM BoI quantities.

Below, we present a simulation-free scheme that compares the BoIs of a given pair of LPM and DPM in terms of the $L^{\infty}$ norm, for the purpose of model consistency analysis.

\section{Simulation-free scheme} \label{sec:scheme}

Assuming that an LPM and a DPM are provided, along with their matching mass properties, ICs/BCs, and source terms, as well as the BoI quantities' matching form (i.e., spatial integration over the boundary surface), a scheme is presented below for calculating the a priori error between the corresponding BoI quantities of the two models, without requiring explicit numerical solution of differential equations.

We start by deriving the governing equations for both the LPM and the DPM, which are respectively described by a system of second-order ODEs and PDEs, respectively. To enable a comparison between the two models, we discretize the PDEs of the DPM using spatial discretization methods \cite{bathe2006finite}, thereby converting them to a system of second-order ODEs. We then transform the given ODEs of the LPM and the resulting ODEs (semi-discretized PDEs) of the DPM into their respective state-space forms through linear transformation \cite{atkinson2011numerical}, as illustrated below:
\begin{itemize}
    \item LPM ODEs in state-space form:
    \begin{align}
        &\begin{array}{*{20}{l}}
            {{{\bf{E}}_l}{{{\bf{\dot x}}}_l}(t)\;{\kern 1pt}  = {{\bf{A}}_l}{{\bf{x}}_l}(t) + {{\bf{B}}_l}{{h}_l}(t)}\\
            {{{\bf{y}}_l}(t) = {{\bf{C}}_l}{{\bf{x}}_l}(t)}
        \end{array}  \label{Eq:LPM}
    \end{align}
    \item DPM ODEs (semi-discretized PDEs) in state-space form:
    \begin{align}
        &\begin{array}{*{20}{l}}
            {{{\bf{E}}_d}{{{\bf{\dot x}}}_d}(t)\;{\kern 1pt}  = {{\bf{A}}_d}{{\bf{x}}_d}(t) + {{\bf{B}}_d}{{h}_d}(t)}\\
            {{{\bf{y}}_d}(t) = {{\bf{C}}_d}{{\bf{x}}_d}(t)}
        \end{array}  \label{Eq:DPM}
    \end{align}
\end{itemize}

A human-readable LPM ODE system (\ref{Eq:LPM}) can have tens to hundreds of state variables, while the DPM ODE system (\ref{Eq:DPM}) can have tens or hundreds of thousands, if not millions, of state variables, as a result of semi-discretizing PDEs on a fine mesh. In both equations, ${{\bf{x}}}(t)$ represents a vector of state variables, ${{\bf{y}}(t)}$ represents a vector of BoIs, and $h(t)$ represents the input signal of external forces. The descriptor matrix $\bf{E}$, dynamic matrix $\bf{A}$, input matrix $\bf{B}$, and output matrix $\bf{C}$ parameterize the LTI systems. The user determines the matrix $\bf{C}$ based on the BoI. The matrices ${\bf{A}}$ through ${\bf{E}}$ have the following forms:

\begin{equation} \label{Eq:ABCDE}
{\bf{A}}=\begin{bmatrix}
 \bf{0} & \bf{I}\\ 
 -\bf{K}&  -\bf{R}
\end{bmatrix},
{\bf{B}}=\begin{bmatrix}
\bf{0}\\ \bf{F}
\end{bmatrix},
{\bf{C}}={\bf{C}},
{\bf{E}}=\begin{bmatrix}
 \bf{I} & \bf{0}\\ 
 \bf{0}&  \bf{M}
\end{bmatrix},
\end{equation}

\noindent where the matrix $\bf{I}$ is the identity matrix, $\bf{M}$, $\bf{K}$, and $\bf{R}$ are the mass, stiffness, and damping matrices, respectively, and $\bf{F}$ is a vector that contains the external forces. The matrices ${\bf{M}}$, ${\bf{K}}$, and ${\bf{R}}$ used in ${\bf{A}}_l$ through ${\bf{E}}_l$ in (\ref{Eq:LPM}) are constructed based on the constitutive relationship of lumped components and their adjacency relations. On the other hand, the matrices ${\bf{M}}$, ${\bf{K}}$, and ${\bf{R}}$ used in ${\bf{A}}_d$ through ${\bf{E}}_d$ in (\ref{Eq:DPM}) are generated from spatial discretization methods \cite{mattiussi2000finite}.

Since both systems of ODEs are LTI, we can match them using linear projections:

\begin{enumerate}[(a)]

\item  The source terms are related by ${{\bf{B}}_l(t)} {{h}_l(t)}= {\Gamma _n}{{\bf{B}}_d(t)}{{h}_d(t)}$, where ${\Gamma _n}$ represents a projection matrix that maps the discrete form of forces of DPM to the lumped forces of LPM. 

\item   The initial displacement and velocity terms are related by ${\Gamma _I}{{\bf{x}}_d}\left( 0 \right) = {{\bf{x}}_l}\left( 0 \right)$ and ${\Gamma _I}{\dot{\bf{x}}}_d(0) = {\dot{\bf{x}}}_l(0)$, where ${\Gamma _I}$ denotes the projection matrix that maps  the spatially discretized ICs of the DPM to the ICs of the LPM.

\item  The BoI terms are related by ${{\bf{C}}_d} = {\Gamma _f}$, where $\Gamma _f$ is a projection matrix that maps the spatially discretized BoI of DPM to the BoI of LPM.

\end{enumerate}

By substituting the projection in (a) into (\ref{Eq:LPM}), we obtain a revised state-space form:

\begin{equation} \label{Eq:LPM_u_updated}
\begin{array}{*{20}{l}}
{{{\bf{E}}_l}{{{\bf{\dot x}}}_l}(t)\;{\kern 1pt}  = {{\bf{A}}_l}{{\bf{x}}_l}(t) + {\bf{B}}_l^\prime{{h}_d}(t)}\\
{{{\bf{y}}_l}(t) = {{\bf{C}}_l}{{\bf{x}}_l}(t)}
\end{array}
\end{equation}
This new form indicates that the source term of the LPM can be replaced with the source term of the DPM. This substitution is crucial for computing an upper-bound for the $L^{\infty}$ error between ${\bf{y}}_l(t)$ and ${\bf{y}}_d(t)$, as demonstrated below.

Given that we are working with LTI systems, we can solve (\ref{Eq:LPM_u_updated}) and (\ref{Eq:DPM}) using Laplace transforms \cite{callier2012linear}. The solutions take the following forms:

\begin{subequations}
\begin{align}
&{{\bf{y}}_l}(s) = \underbrace {\left( {{{\bf{C}}_l}{{\left( {s{{\bf{E}}_l} - {{\bf{A}}_l}} \right)}^{ - 1}}{{{\bf{B'}}}_l}} \right)}_{{{\bf{G}}_l}(s)}{{h}_d}(s) \label{Eq:soln_LPM_u_updated}
\\
&{{\bf{y}}_d}(s) = \underbrace {\left( {{{\bf{C}}_d}{{\left( {s{{\bf{E}}_d} - {{\bf{A}}_d}} \right)}^{ - 1}}{{\bf{B}}_d} } \right)}_{{{\bf{G}}_d}(s)}{{h}_d}(s), \label{Eq:soln_DPM}
\end{align}
\end{subequations}

\noindent where $s$ represents the complex frequency variable, while ${{{\bf{G}}_l}(s)}$ and ${{{\bf{G}}_d}(s)}$ correspond to the transfer functions of the LPM and semi-discretized DPM, respectively.

When two models with the same input ${h}_d(t)$ are compared, the maximum difference between their outputs ${{{\bf{y}}_d}(t)}$ and ${{{\bf{y}}_l}(t)}$ is upper-bounded \cite{antoulas2010interpolatory} as follows:

\begin{equation} \label{Eq:bound}
\mathop {\max }\limits_{t \in [0, \infty)} {\left\| {{{\bf{y}}_d}(t) - {{\bf{y}}_l}(t)} \right\|_\infty } \le {\left\| {{{\bf{G}}_d}(s) - {{\bf{G}}_l}(s)} \right\|_{{\mathcal{H}_2}}} \cdot \sqrt {\int_0^\infty  {{{\left\| {{{h}_d}(t)} \right\|}_2}dt}},
\end{equation}
\noindent as long as ${{h}_d(t)}$ is a finite energy input (i.e., $\int_0^\infty {\left\| {{{h}_d}(t)} \right\|_2^2dt < \infty } $), which is the case for most engineering problems. The formula to compute ${\left\| {{{\bf{G}}_d}(s) - {{\bf{G}}_l}(s)} \right\|_{{\mathcal{H}_2}}}$ is given in \ref{Eq:H2norm}, where j is the imaginary unit and $\omega$ is the frequency in radians per unit time.

\begin{equation} \label{Eq:H2norm}
{\left\| {{{\bf{G}}_d}(s) - {{\bf{G}}_l}(s)} \right\|_{{\mathcal{H}_2}}}= \left ( \frac{1}{2\pi}\int_{-\infty }^{\infty}\left | {\bf{G}}_d(j\omega)- {\bf{G}}_l(j\omega) \right |d\omega   \right )^{1/2}
\end{equation}

Notably, the integration term in \ref{Eq:bound} is constant for a given source term ${h}_d(t)$, as the BC of DPM is time-invariant. This means that as the norm ${\left\| {{{\bf{G}}_d}(s) - {{\bf{G}}_l}(s)} \right\|_{{\mathcal{H}_2}}}$ approaches zero, the maximum difference between ${{\bf{y}}_d}(t)$ and ${{\bf{y}}_l}(t)$ also approaches zero.

In essence, the deviation between the BoIs of LPM and DPM in the time domain can be approximated by the deviation between ${{\bf{G}}_d}(s)$ and ${{\bf{G}}_l}(s)$ in the frequency domain using the ${{\mathcal{H}_2}}$ norm. An important observation is that the computation of ${\left\| {{{\bf{G}}_d}(s) - {{\bf{G}}_l}(s)} \right\|_{{\mathcal{H}_2}}}$ involves solving algebraic equations instead of ODEs, as discussed in \cite{peeters2013computing}. This means that we can avoid solving ODEs altogether. To speed up the computation of these algebraic equations, we use the ROM of the DPM, as shown in the equation below.

\begin{equation} \label{Eq:LPM_Mr}
\begin{array}{*{20}{l}}
{{{\bf{E}}_r}{{{\bf{\dot x}}}_r}(t)\;{\kern 1pt}  = {{\bf{A}}_r}{{\bf{x}}_r}(t) + {{\bf{B}}_r}{{h}_d}(t)}\\
{{{\bf{y}}_r}(t) = {{\bf{C}}_r}{{\bf{x}}_r}(t)}
\end{array}
\end{equation}

In order to enhance the computational efficiency of solving large-scale algebraic equations arising from DPMs, a MOR method is employed. Specifically, we adopt the SPARK+CURE method from \cite{panzer2014model}, which generates a small-scale surrogate model for the DPM. Due to space constraints, please refer to Panzer's original Ph.D. thesis \cite{panzer2014model} for more details about this method. This surrogate model has the same source terms ${{\bf{u}}_d}(t)$ as the DPM in (\ref{Eq:DPM}), and its transfer function is denoted by ${{\bf{G}}}_r(s)$. The SPARK+CURE method guarantees an upper-bound ${\left\| {{{\bf{G}}_d}(s) - {{\bf{G}}_l}(s)} \right\|_{{\mathcal{H}_2}}} \le {{\bar \varepsilon }_1}$ between the transfer functions of the DPM and the surrogate model. Unlike the spatial-discretized DPM, we assume that the given LPM in (\ref{Eq:LPM}) has a small scale, so we do not apply MOR to it.

Because both the given LPM and the surrogate model of the semi-discretized DPM have a small state space, computing the value ${{\bar \varepsilon }_2}$ of ${\left\| {{{\bf{G}}_r}(s) - {{\bf{G}}_l}(s)} \right\|_{{\mathcal{H}_2}}}$ is time-efficient. With these two values, we can use a triangular inequality to obtain an upper-bound  ${\bar \varepsilon }$ for the error between the transfer functions ${{\bf{G}}_d}(s)$ of the DPM and ${{\bf{G}}_l}(s)$ of the LPM as follows:

\begin{equation}  \label{Eq:inequality}
\begin{array}{*{20}{l}}
{{{\left\| {{{\bf{G}}_d}(s) - {{\bf{G}}_l}(s)} \right\|}_{{\mathcal{H}_2}}} = {{\left\| {{{\bf{G}}_d}(s) - {{\bf{G}}_r}(s) + {{\bf{G}}_r}(s) - {{\bf{G}}_l}(s)} \right\|}_{{\mathcal{H}_2}}}}\\
{\;{\kern 1pt} \;{\kern 1pt} \;{\kern 1pt} \;{\kern 1pt} \;{\kern 1pt} \;{\kern 1pt} \;{\kern 1pt} \;{\kern 1pt} \;{\kern 1pt} \;{\kern 1pt} \;{\kern 1pt} \;{\kern 1pt}  \le {{\left\| {{{\bf{G}}_d}(s) - {{\bf{G}}_r}(s)} \right\|}_{{\mathcal{H}_2}}} + {{\left\| {{{\bf{G}}_r}(s) - {{\bf{G}}_l}(s)} \right\|}_{{\mathcal{H}_2}}}}\\
{\;{\kern 1pt} \;{\kern 1pt} \;{\kern 1pt} \;{\kern 1pt} \;{\kern 1pt} \;{\kern 1pt} \;{\kern 1pt} \;{\kern 1pt} \;{\kern 1pt} \;{\kern 1pt} \;{\kern 1pt} \;{\kern 1pt}  \le {{\bar \varepsilon }_1} + {{\bar \varepsilon }_2} = {{\bar \varepsilon }}}
\end{array}
\end{equation}
If the upper-bound ${\bar \varepsilon }$ is within the given tolerance, we can view the BoI of DPM and the BoI of LPM as similar.

The equation above provides an upper-bound for the absolute error. To obtain an upper-bound for the relative error, we can use the equation below, where the transfer function ${{\bf{G}}_l}(s)$ of the LPM is selected as the reference.

\begin{equation}  \label{Eq:inequality_rel}
{{{{{\left\| {{{\bf{G}}_d}(s) - {{\bf{G}}_l}(s)} \right\|}_{{{\cal H}_2}}}} \over {{{\left\| {{{\bf{G}}_l}(s)} \right\|}_{{{\cal H}_2}}}}} \le {{{{\bar \varepsilon }_1} + {{\bar \varepsilon }_2}} \over {{{\left\| {{{\bf{G}}_l}(s)} \right\|}_{{{\mathcal{H}}_2}}}}} = {{\bar \varepsilon }_{rel}}}
\end{equation}

    \begin{figure}[!htb]
        \centering
        \includegraphics[width=0.8\linewidth]{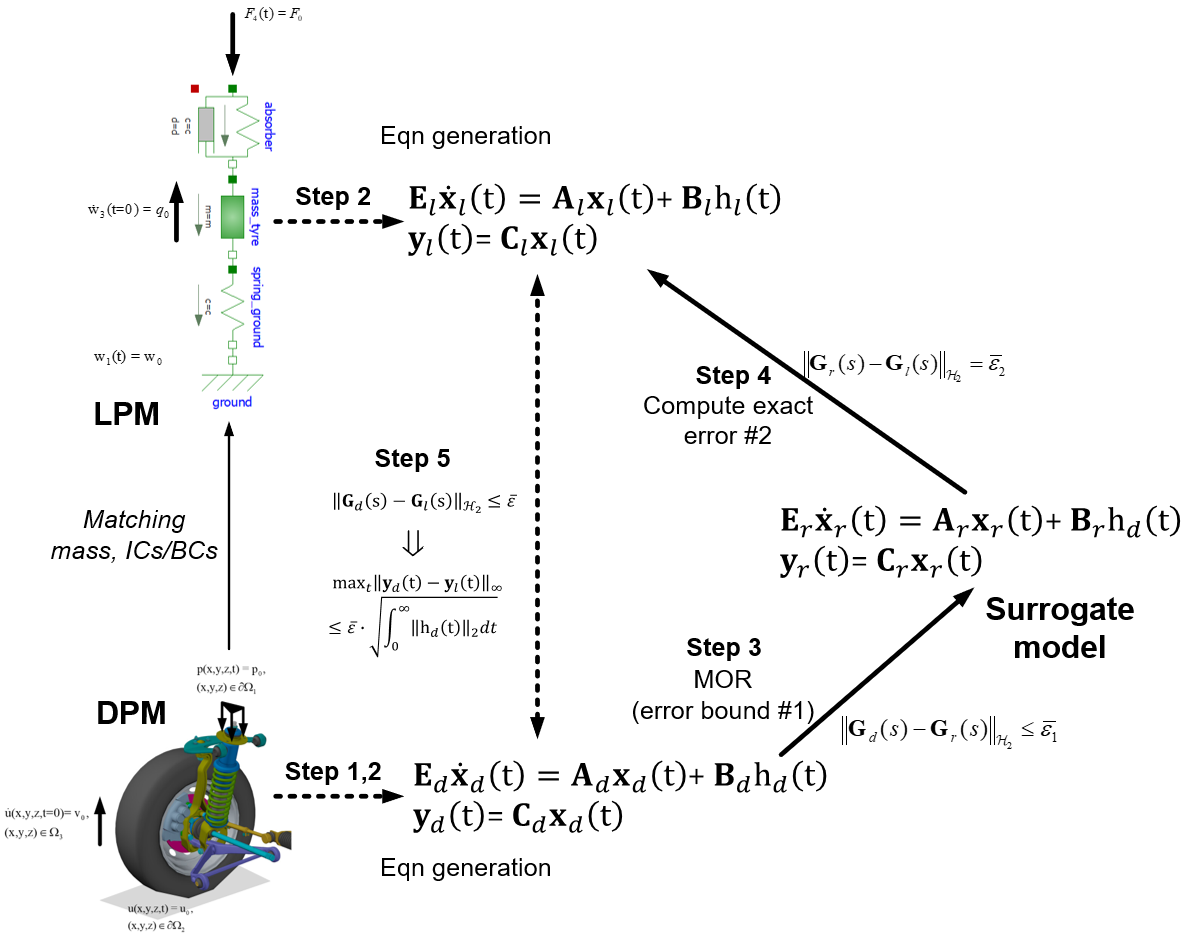}
        \caption{Simulation-free scheme to compute the error bound between the BoIs of a pair of LPM and DPM}
        \label{fig:4steps}
    \end{figure}

Figure \ref{fig:4steps} provides a visual representation of the five-step simulation-free scheme proposed in this study, where an LPM with masses, springs, and dampers and a DPM with 3D suspension are used as examples. The  approach is outlined step-by-step as follows. Given the ODEs of the LPM and the PDEs of the DPM:

\begin{enumerate}

\item Use a spatial discretization (e.g. finite element \cite{bathe2006finite}) method to convert the PDEs into a system of ODEs. 

\item Convert the ODEs of both models to state-space form.

\item Use the SPARK+CURE method to reduce the scale of the DPM and ensure a priori guaranteed $\mathcal{H}_2$ error ${{{\bar \varepsilon }_1}}$ between the DPM and its surrogate model.

\item Calculate the exact $\mathcal{H}_2$ error ${{{\bar \varepsilon }_2}}$ between the surrogate DPM and the LPM.

\item Estimate the upper-bound of the $\mathcal{H}_2$ error ${{{\bar \varepsilon }}}$ between the transfer functions of the given LPM and semi-discretized DPM using the triangular inequality and verify whether it falls within the acceptable tolerance.

\end{enumerate}

\section{Applications}  \label{sec:application}
In this section, we demonstrate the practical application of the proposed simulation-free scheme by using it to compare the solutions of the LPM and the DPM for two mechanical designs: a bracket (Figure \ref{fig:bracket}) and a frame (Figure \ref{fig:bike}). Both designs have linear isotropic material properties, and their mass properties, ICs/BCs, and the form of the BoI quantities are pre-specified to ensure consistency between the two models.

\subsection{Model Specifications}
\begin{figure}[!htb]
        \centering
        \begin{minipage}{0.2236\textwidth} % 0.172
        \centering
        \includegraphics[width=\linewidth]{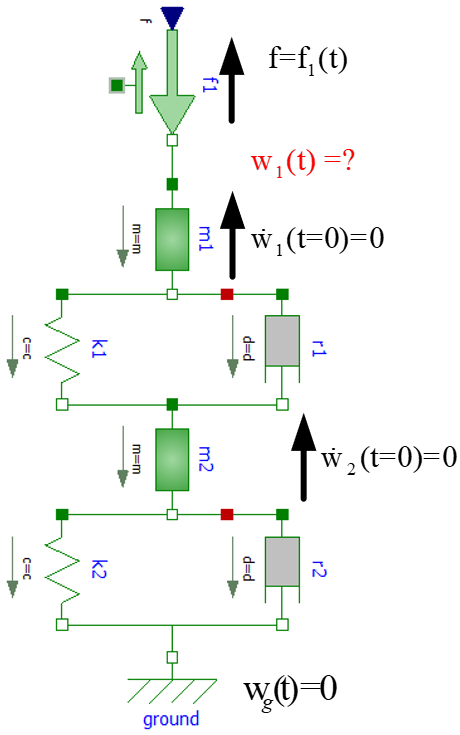} 
        \subcaption{LPM \label{fig:LPM_bracket_bike}}        
        \end{minipage} 
 	\begin{minipage}{0.26\textwidth} % 0.2
        \centering
        \includegraphics[width=\linewidth]{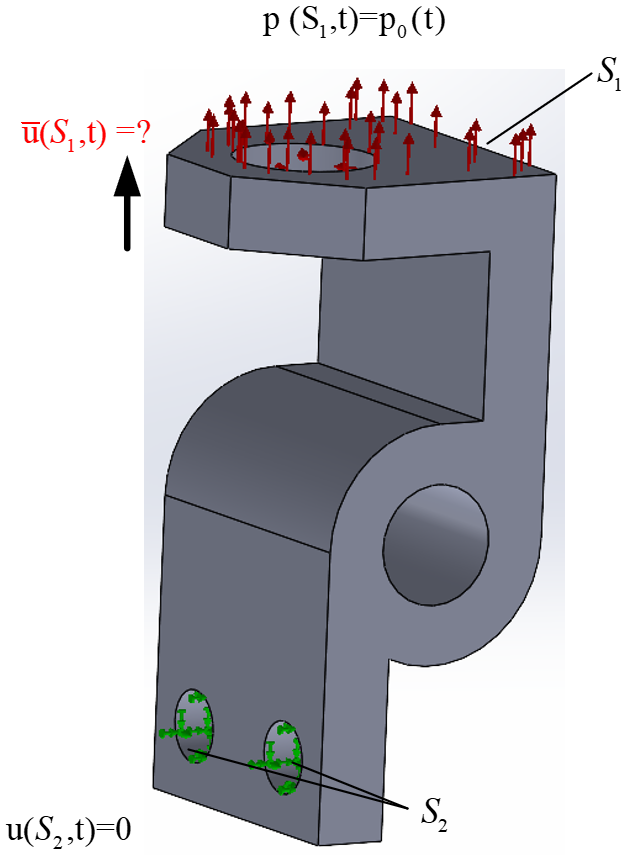} 
        \subcaption{DPM of the bracket \label{fig:bracket}}        
        \end{minipage} 
        \begin{minipage}{0.468\textwidth} % 0.36
        \centering
        \includegraphics[width=\linewidth]{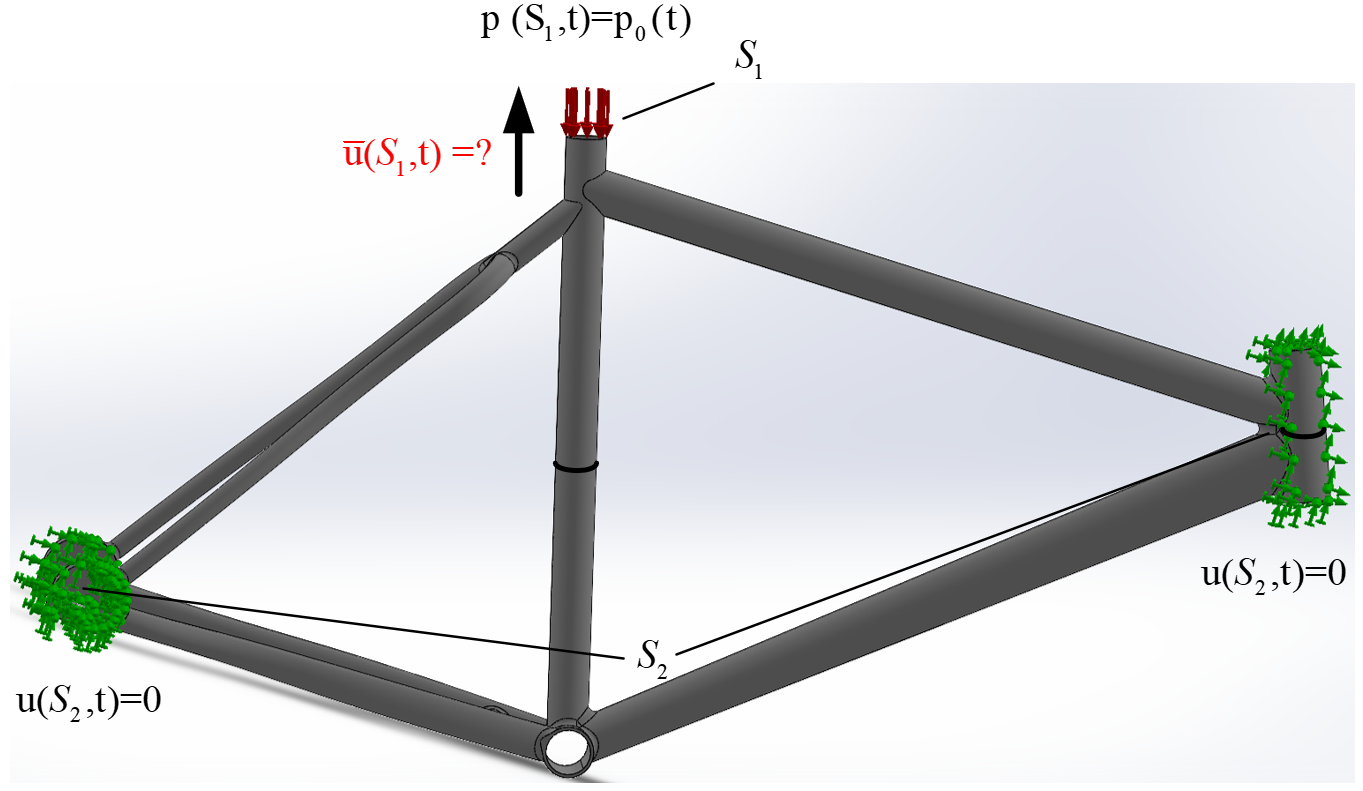}
        \subcaption{DPM of the frame\label{fig:bike}}                 
        \end{minipage} 
        \caption{LPM and DPMs of bracket and frame \label{fig:DPM_bracket_bike}}        
\end{figure}

The same LPM topology is utilized for both mechanical designs, as shown in Figure \ref{fig:LPM_bracket_bike}, but with different parameters for the lumped components. The values are provided in Table \ref{Table:data_LPM}, using SI base units. Both lumped masses have zero initial displacements and velocities.

\begin{table*}[ht] 
\captionof{table} {Parameters of the LPM used for the bracket and the frame designs \label{Table:data_LPM}}
\centering
\resizebox{\textwidth}{!}{ \begin{tabular}{|c|c|c|c|c|c|c|c|}
\hline
           & $m_1$ & $m_2$ & $k_1$ & $k_2$ & $r_1$ & $r_2$ & $f_1$ \\ \hline
LPM for Bracket    & $3.8465 \times {10^5}$  & $3.512 \times {10^3}$  & $3.316 \times {10^4}$  & $4.688\times {10^3}$  & $1.4697\times {10^5}$  & $2.9052 \times {10^3}$  & $0.005$  \\ \hline
LPM for Frame & $7.997 \times {10^5}$  & $6.9139 \times {10^4}$  & $4.8561 \times {10^8}$  & $2.2308\times {10^8}$  & $2.8102\times {10^7}$  & $1.5075 \times {10^5}$  & $2186.56$  \\ \hline
\end{tabular} }
\end{table*}

\begin{figure}[!htb]
        \centering
		\begin{minipage}{0.3887\textwidth}
          \centering
          \includegraphics[width=\linewidth]{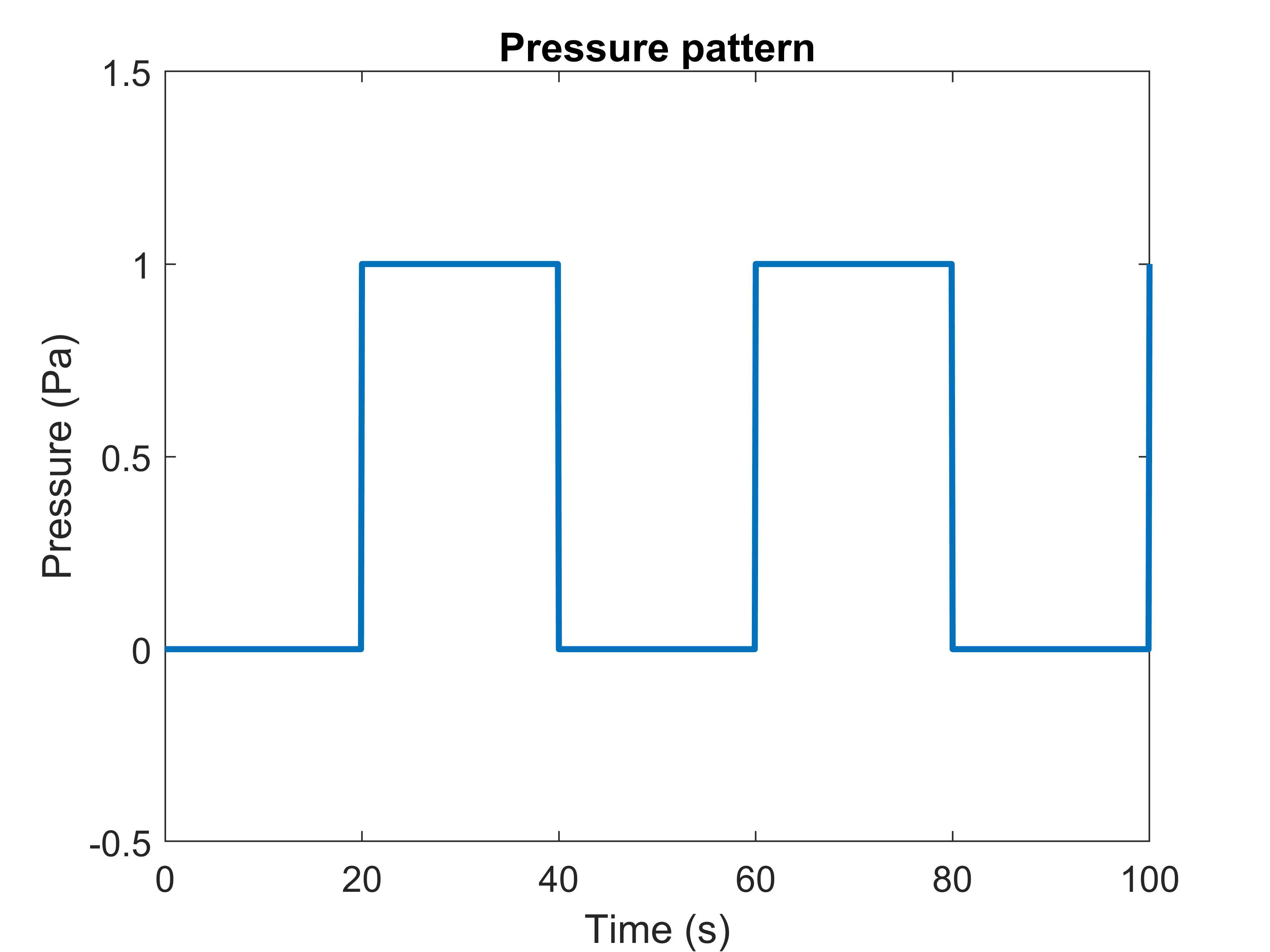}
          \subcaption{Pressure on bracket \label{fig:bracket_pressure}}                 
        \end{minipage} 
 		\begin{minipage}{0.3887\textwidth}
        \centering
        \includegraphics[width=\linewidth]{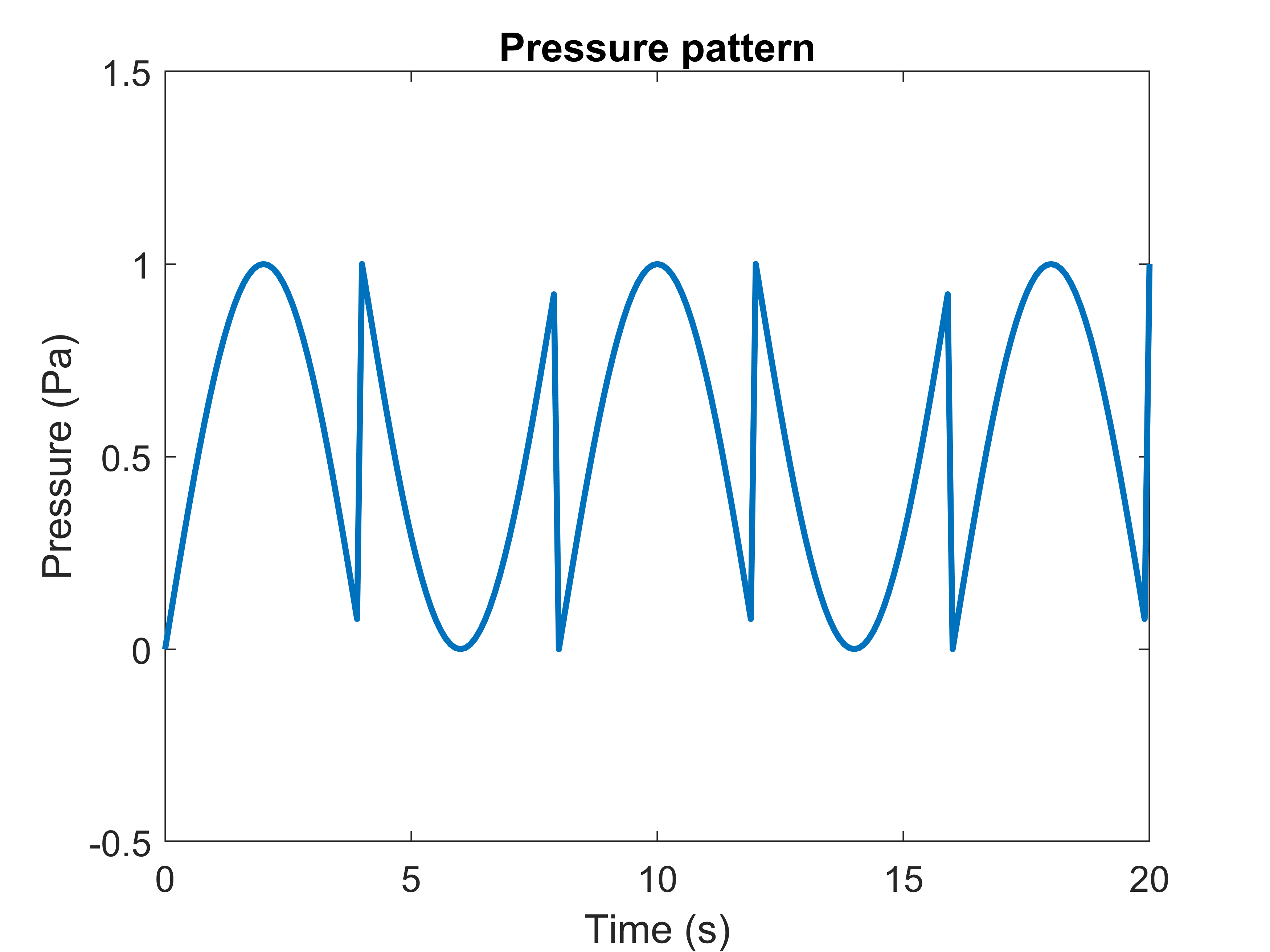} 
        \subcaption{Pressure on frame \label{fig:bike_pressure}}        
        \end{minipage}        
        \caption{Pressures on bracket and frame \label{fig:BC_bracket_bike}}        
\end{figure}

Two mechanical parts, namely a bracket and a frame, have been designed to implement the LPM instances. The DPMs for both designs have zero initial displacements and velocities. The top surfaces of the bracket and the frame are subjected to pressures that vary with time. The time-varying characteristics of the pressures are depicted in Figures \ref{fig:bracket_pressure} and \ref{fig:bike_pressure} for the bracket and the frame, respectively.

In order to compare the LPM and DPM models of the two designs, it is necessary to ensure that their mass properties, ICs/BCs, and BoI quantities match. The requirements for this matching are explained in (C1)-(C3) in Section \ref{sec:scheme}, and an example of the matching process is provided in Section \ref{sec:a_me_ex}. To maintain consistency, we will follow the same matching procedure as the example in Section \ref{sec:a_me_ex}.

\begin{figure}[!htb]
        \centering
		\begin{minipage}{0.2873\textwidth}
          \centering
          \includegraphics[width=0.5\linewidth]{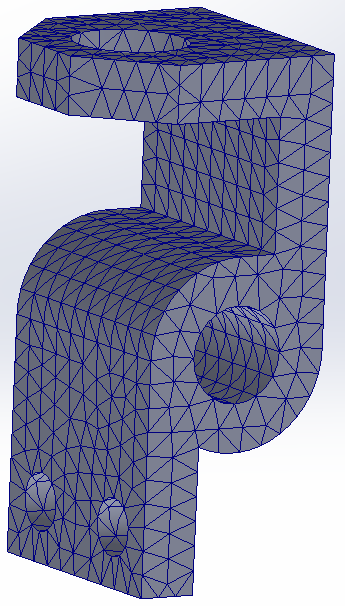}
          \subcaption{Mesh of bracket \label{fig:mesh_bracket}}                 
        \end{minipage} 
 		\begin{minipage}{0.507\textwidth}
        \centering
        \includegraphics[width=\linewidth]{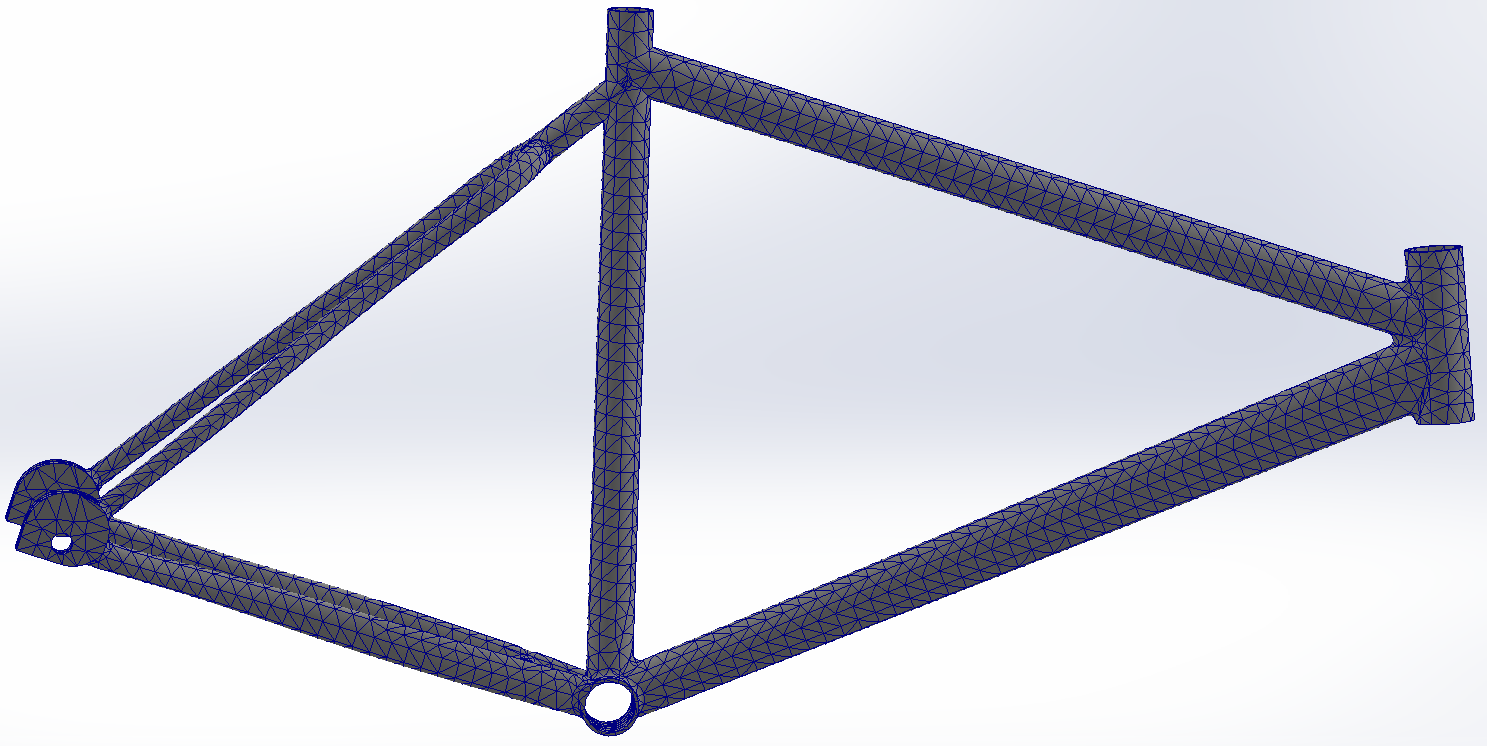} 
        \subcaption{Mesh of frame \label{fig:mesh_bike}}        
        \end{minipage}        
        \caption{Meshes of bracket and frame \label{fig:mesh_bracket_bike}}        
\end{figure} 

\subsection{Surrogate DPM of the Bracket}
Here, we demonstrate the effectiveness of our simulation-free approach in enhancing the efficiency of model comparison for large-scale models. We used tetrahedral finite elements to discretize the bracket's geometry at three different resolutions (only one sketch is shown in Figure \ref{fig:behavior_bracket_case1}), resulting in three sets of state-space equations with a state variable count ranging from approximately 3,000 to 6,000. We then applied the SPARK+CURE method to generate three sets of ROMs that have a priori relative ${{\mathcal{H}}_2}$  error bounds just below 1\%.

To assess the quality of the ROMs, we used the backward Euler method \cite{mazumder2015numerical} to simulate the DPM and all its surrogate models, using the same solver and time steps. The simulation results are presented in Figures \ref{fig:behavior_bracket_case1} through \ref{fig:behavior_bracket_case3}, where the abscissa denotes computation time and the ordinate denotes the average displacement of the bracket's top surface. Due to limited space, only a few ROM curves are labeled. The solver employed in the simulations was the ``sparse state space (sss)'' toolbox in Matlab \cite{castagnotto2017sss}, which can analyze high-dimensional dynamical systems with state-space dimensions of $O(10^4)$ or more. This solver preserves system matrix sparsity, allowing it to use computationally efficient operations, such as sparse LU decompositions \cite{trefethen2022numerical}, that would otherwise be infeasible or time-consuming. It is important to note that the simulations were only used to illustrate the accuracy of the surrogate models and were not necessary for our proposed model comparison method.

\subsection{Error Analysis}

\begin{figure}[!htb]
	\begin{minipage}{0.46\textwidth}
		\centering
		\includegraphics[trim = 3.7cm 8cm 4.5cm 8cm,clip ,width=\linewidth]{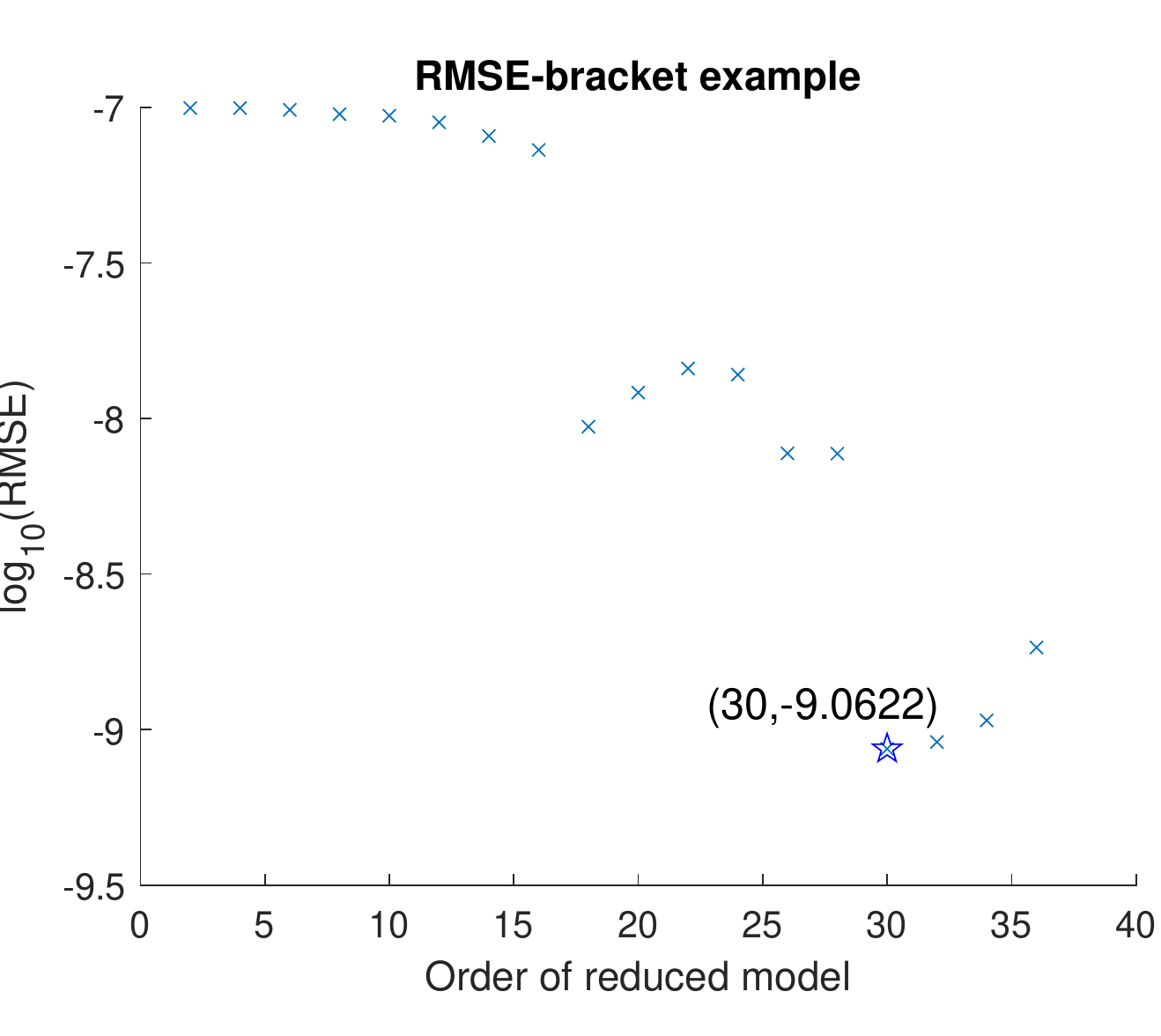}
		\subcaption{RMSE of Case 1 \label{fig:RMSE_bracket_case1}}		
	\end{minipage}  
	\begin{minipage}{0.46\textwidth}
		\centering
		\includegraphics[trim = 3.6cm 8cm 4.6cm 8cm,clip ,width=\linewidth]{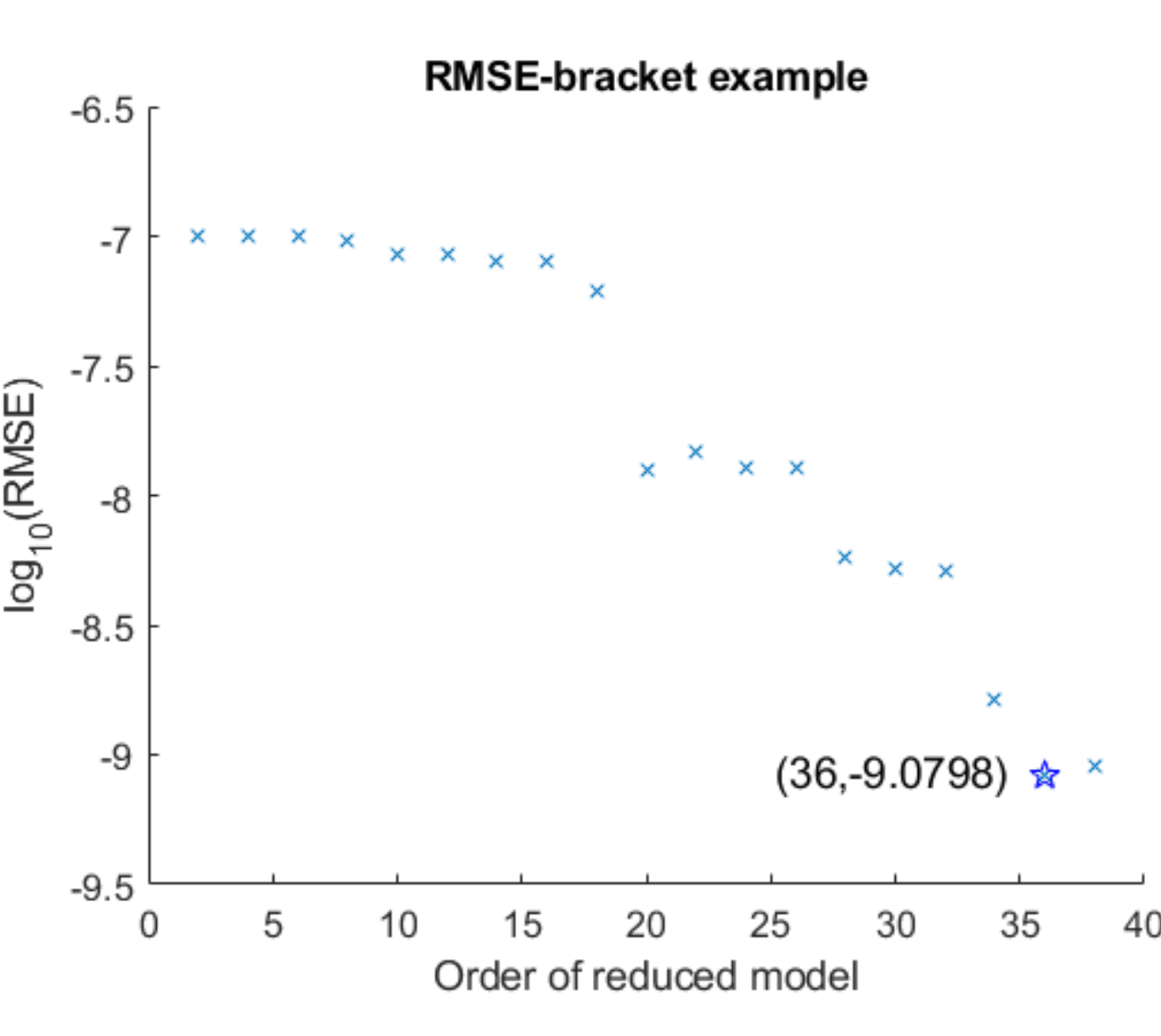}
		\subcaption{RMSE of Case 2\label{fig:RMSE_bracket_case2}}		
	\end{minipage}        
	\begin{minipage}{0.46\textwidth}
		\centering
		\includegraphics[trim = 3.7cm 8cm 4.5cm 8cm,clip ,width=\linewidth]{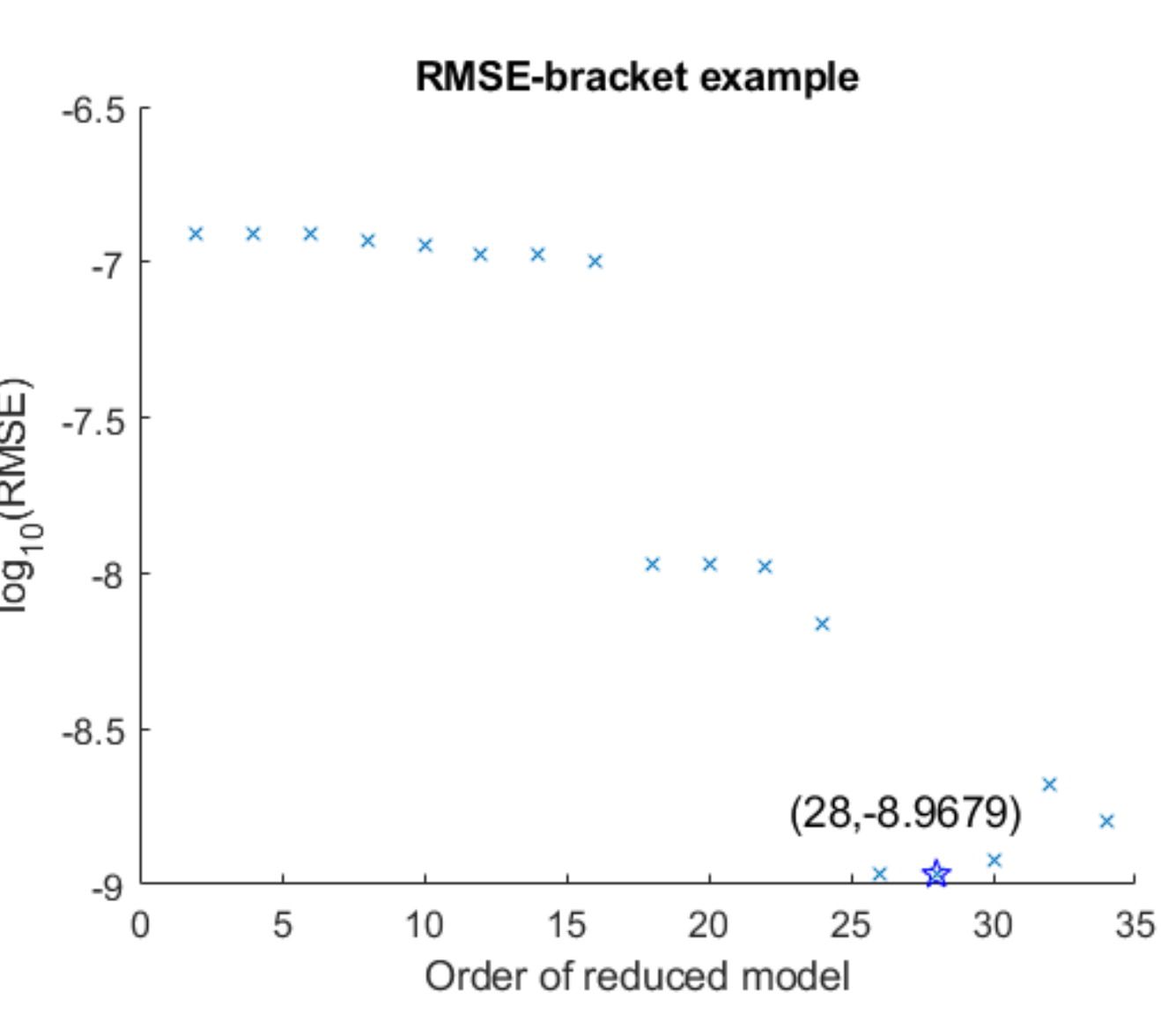}
		\subcaption{RMSE of Case 3 \label{fig:RMSE_bracket_case3}}		
	\end{minipage}
	\begin{minipage}{0.46\textwidth}
		\centering
		\includegraphics[trim = 3.5cm 8cm 4.6cm 8cm,clip ,width=\linewidth]{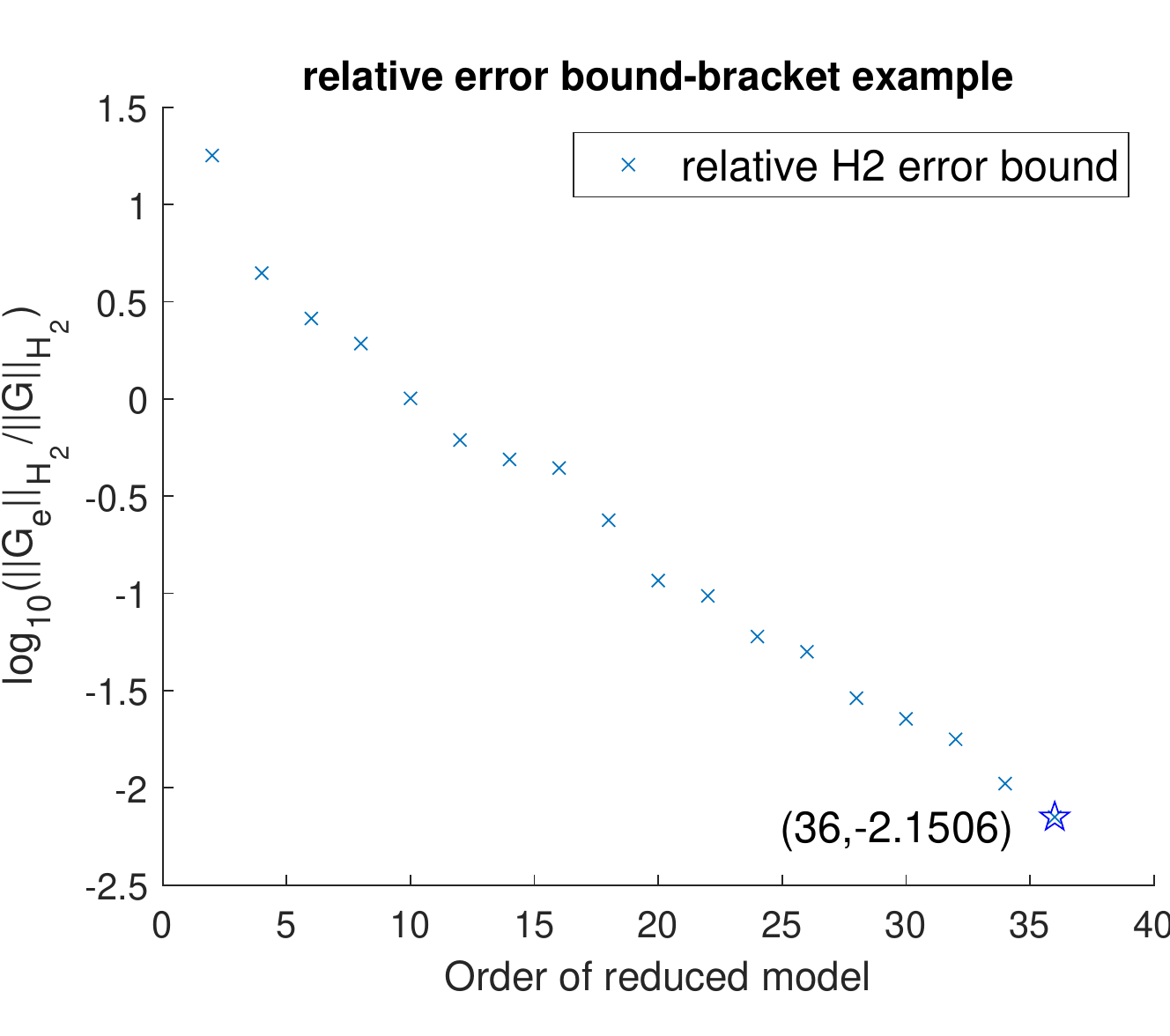}
		\subcaption{Error bound of Case 1 \label{fig:error_bound_bracket_case1}}		
	\end{minipage}  
	\begin{minipage}{0.46\textwidth}
		\centering
		\includegraphics[trim = 3.5cm 8cm 4.55cm 8cm,clip ,width=\linewidth]{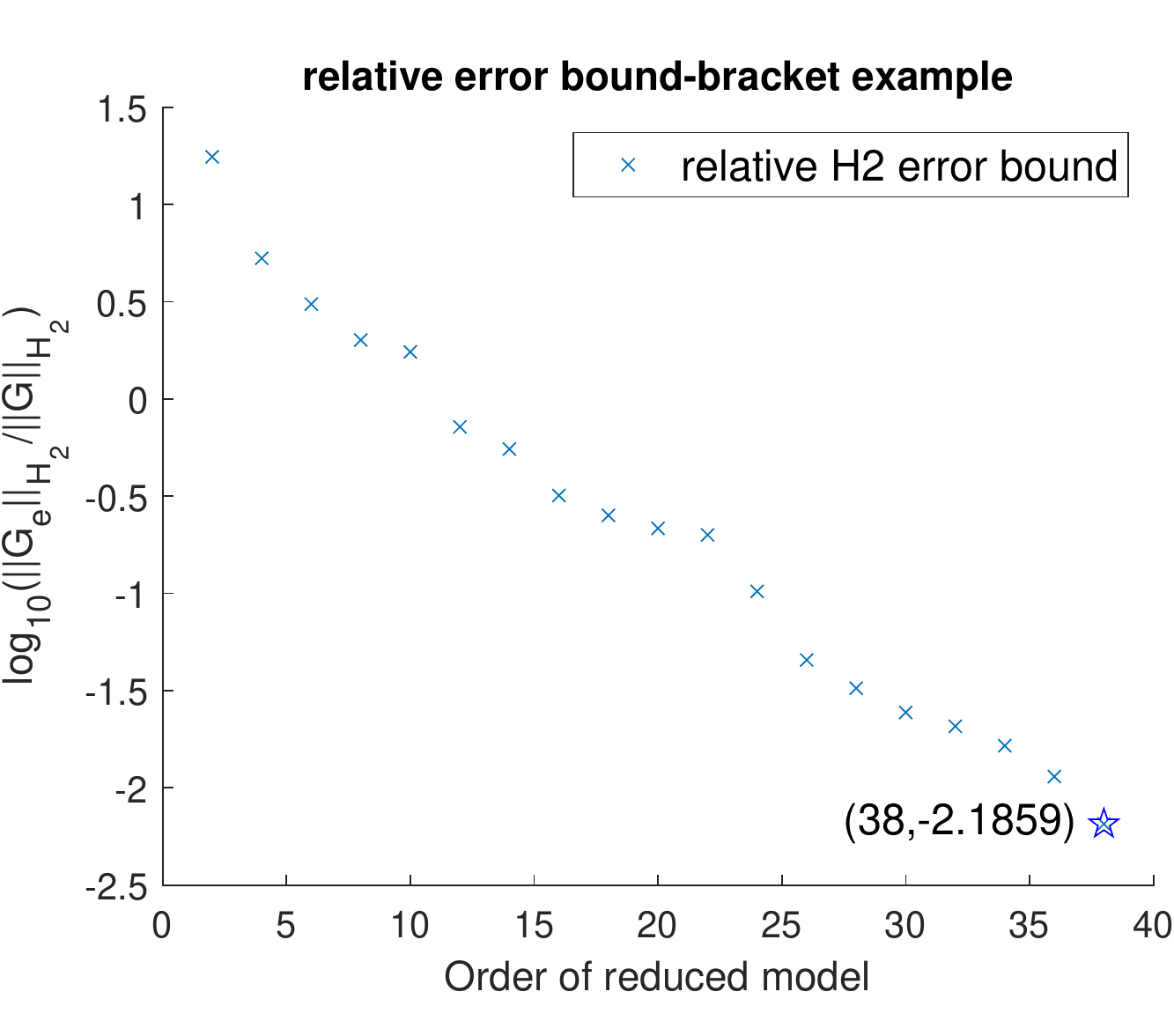}
		\subcaption{Error bound of Case 2\label{fig:error_bound_bracket_case2}}		
	\end{minipage}        
	\begin{minipage}{0.46\textwidth}
		\centering
		\includegraphics[trim = 3.5cm 8cm 4.55cm 8cm,clip ,width=\linewidth]{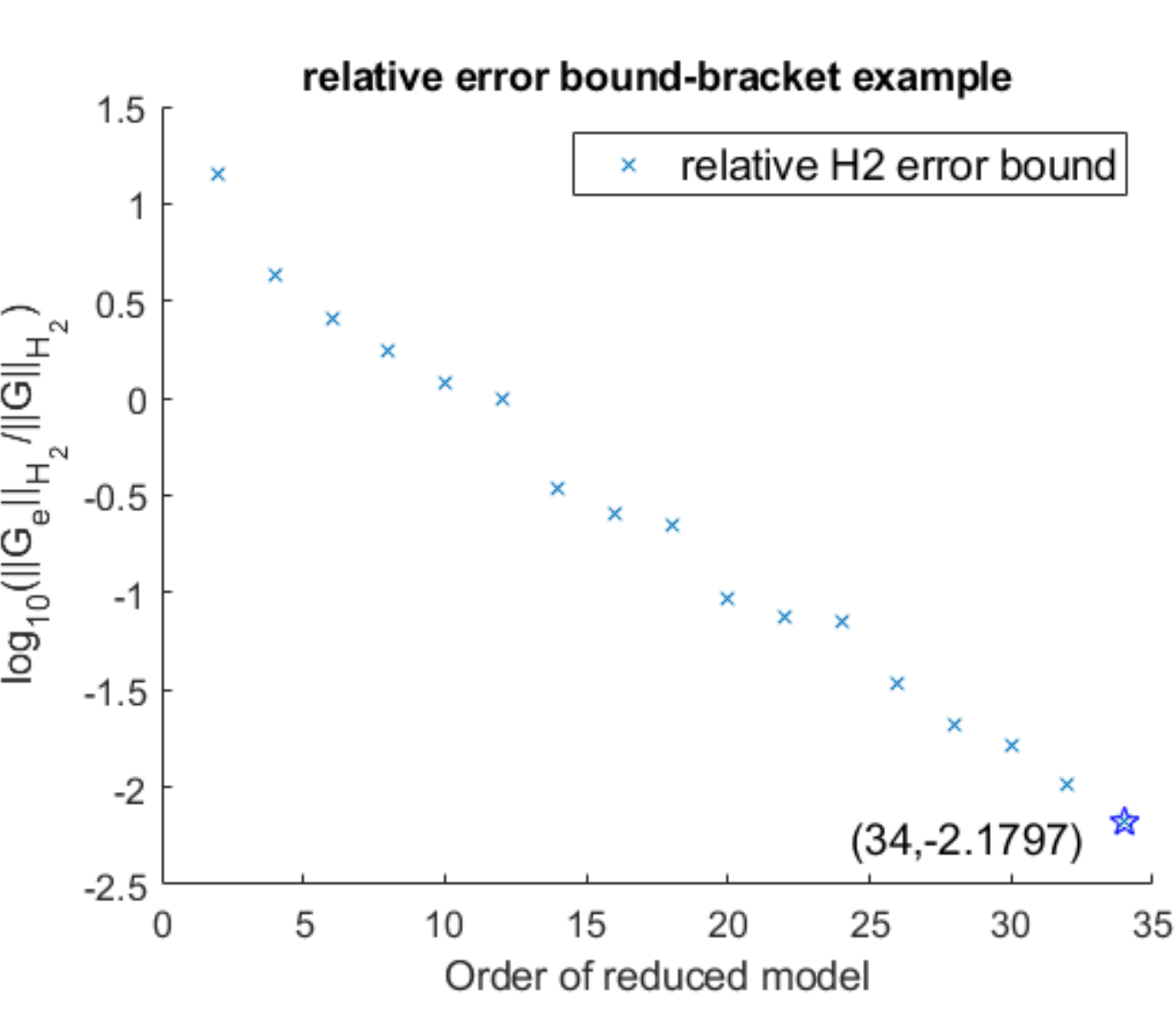}
		\subcaption{Error bound of Case 3 \label{fig:error_bound_bracket_case3}}		
	\end{minipage}
		\caption{RMSE and a priori relative ${{\mathcal{H}}_2}$ error bound - the bracket example\label{fig:RMSE_error_bound_bracket}}         
\end{figure}

As the order of ROM increased, which corresponds to an increase in the number of state variables of ROM, the simulated response of the ROM more closely approached that of the DPM for all cases. To quantify the error introduced by the MOR process, we calculated two different measures, as shown in Figure \ref{fig:RMSE_error_bound_bracket}: the relative $\mathcal{H}_2$ error of the transfer functions and the root mean square error (RMSE) of the simulated response, whose formulas can be found in \cite{wang2021consistency}. The SPARK+CURE method guaranteed the strictly monotonic decay of the bound of the a priori relative $\mathcal{H}_2$ error, as shown in Figures \ref{fig:error_bound_bracket_case1} $\sim$ \ref{fig:error_bound_bracket_case3}. However, it did not provide a rigorous a priori bound on RMSE, as shown in Figures \ref{fig:RMSE_bracket_case1} $\sim$ \ref{fig:RMSE_bracket_case3}, although an overall reduction pattern was commonly observed.

For all three cases, the RMSE was around $O({10^{ - 9}})$, two orders of magnitude smaller than the DPM output of $O({10^{ - 7}})$. The a priori relative $\mathcal{H}_2$ error bound showed that if the order of ROM was larger than 36, 38, and 34 in the three cases, respectively, the exact relative $\mathcal{H}_2$ error would be less than ${10^{ - 2.1506}} \approx 0.707\%$, ${10^{ - 2.1859}} \approx 0.652\%$ and ${10^{ - 2.1797}} \approx 0.661\%$. By selecting the ROM of order 34 generated from case 3 as the surrogate DPM for the bracket, we computed the a priori relative $\mathcal{H}_2$ error bound between the discretized DPM and the LPM using (\ref{Eq:inequality_rel}) as follows:

\begin{equation}
\begin{gathered}
  \frac{{{{\left\| {{{\mathbf{G}}_d}(s) - {{\mathbf{G}}_l}(s)} \right\|}_{{\mathcal{H}_2}}}}}{{{{\left\| {{{\mathbf{G}}_l}(s)} \right\|}_{{\mathcal{H}_2}}}}} \le {\text{ }}\frac{{{{\bar \varepsilon }_1} + {{\bar \varepsilon }_2}}}{{{{\left\| {{{\mathbf{G}}_l}(s)} \right\|}_{{\mathcal{H}_2}}}}}{\text{ }} = {\text{0.0463}} = {\bar \varepsilon }_{rel}
\end{gathered} 
\end{equation}

\begin{figure}[!htb]
	\begin{minipage}{0.46\textwidth}
		\centering
		\includegraphics[trim = 3.8cm 8cm 4.45cm 8cm,clip ,width=\linewidth]{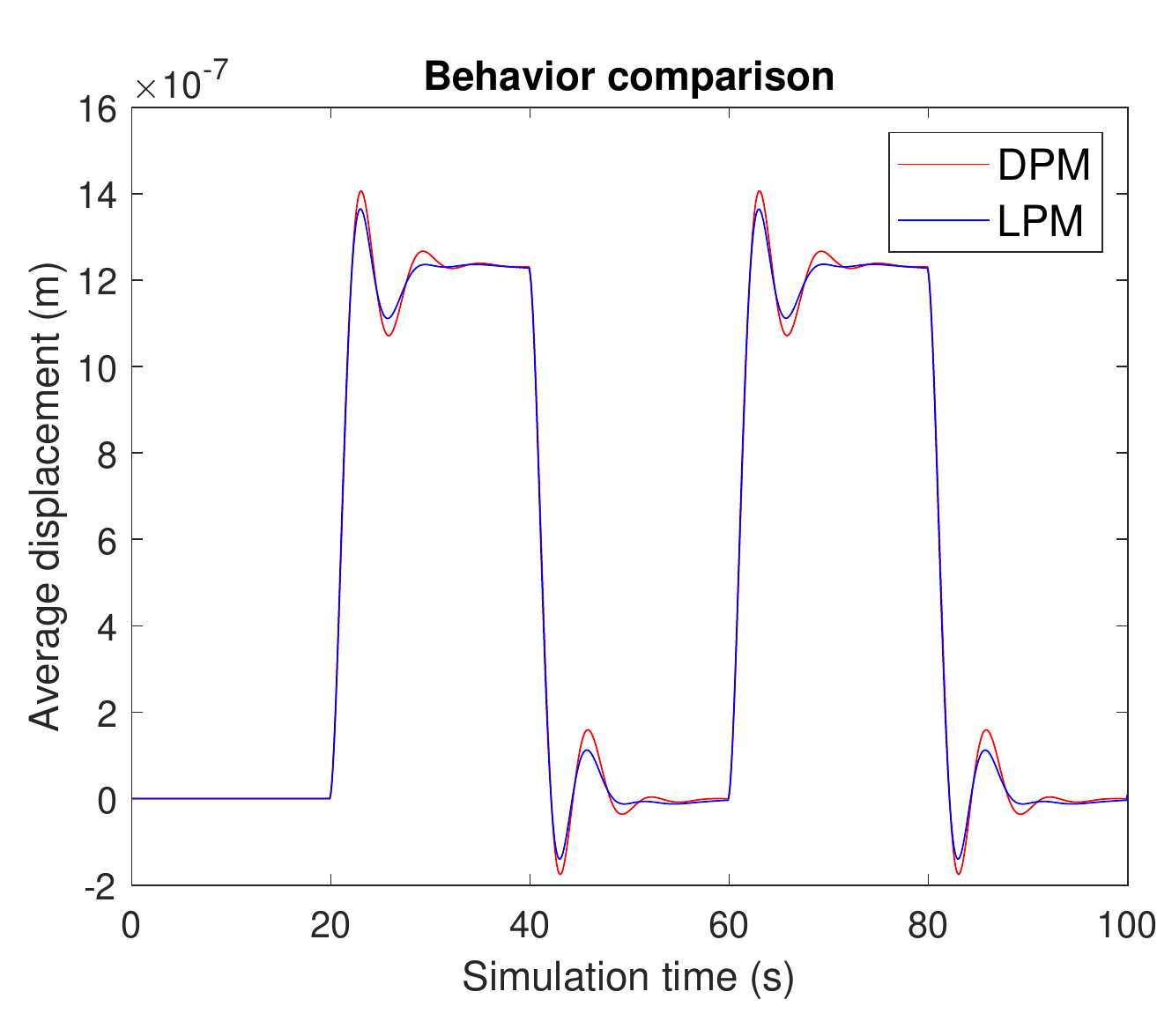}
		\subcaption{Solution comparison \label{fig:comparison_final_bracket}}		
	\end{minipage}  
	\begin{minipage}{0.46\textwidth}
		\centering
		\includegraphics[trim = 3.6cm 8cm 4.45cm 8cm,clip ,width=\linewidth]{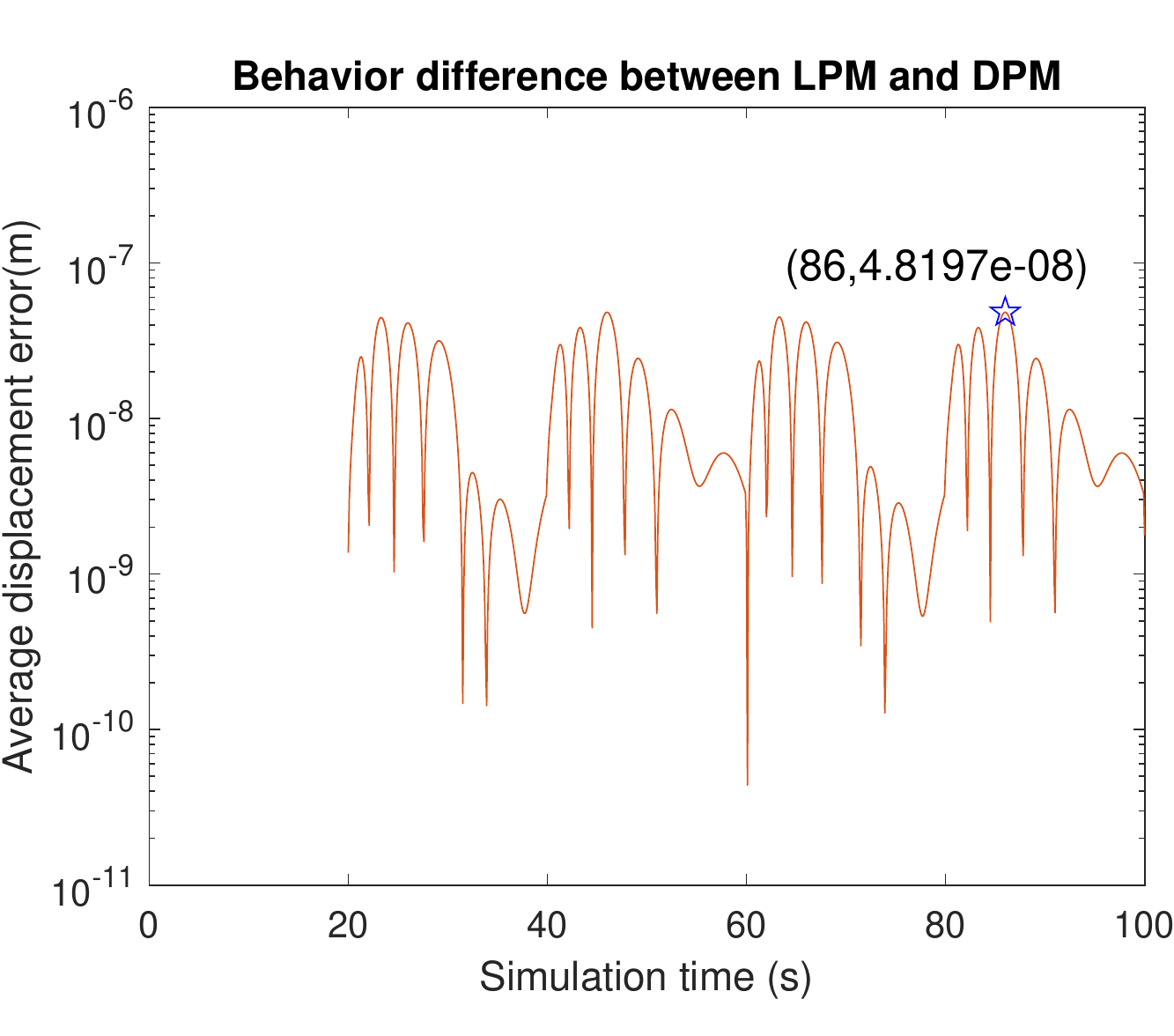}
		\subcaption{Error\label{fig:bound_final_bracket}}	
	\end{minipage}                
	\caption{Numerical solution comparison between the LPM and the DPM - bracket example\label{fig:results_bracket}}	
\end{figure}

Given that the pre-conditions ensure the matches of masses, ICs, and BCs, the LPM and DPM would be considered consistent if the upper bound of error $\bar \varepsilon = 0.0463$ is within the user-defined tolerance. To compare the numerical solutions between the LPM and DPM, we plot the errors in Figure \ref{fig:bound_final_bracket}, while the numerical solutions are compared in Figure \ref{fig:comparison_final_bracket}. The maximum difference between the model solutions over the entire time domain is less than $4.8197 \times {10^{-8}}$, which is one order of magnitude smaller than the maximum model output O($10^{-6}$).

\subsection{Time Efficiency Analysis}
\begin{figure}[!htb]
	\begin{minipage}{0.46\textwidth}
		\centering
		\includegraphics[trim = 4cm 8cm 4.45cm 8cm,clip ,width=\linewidth]{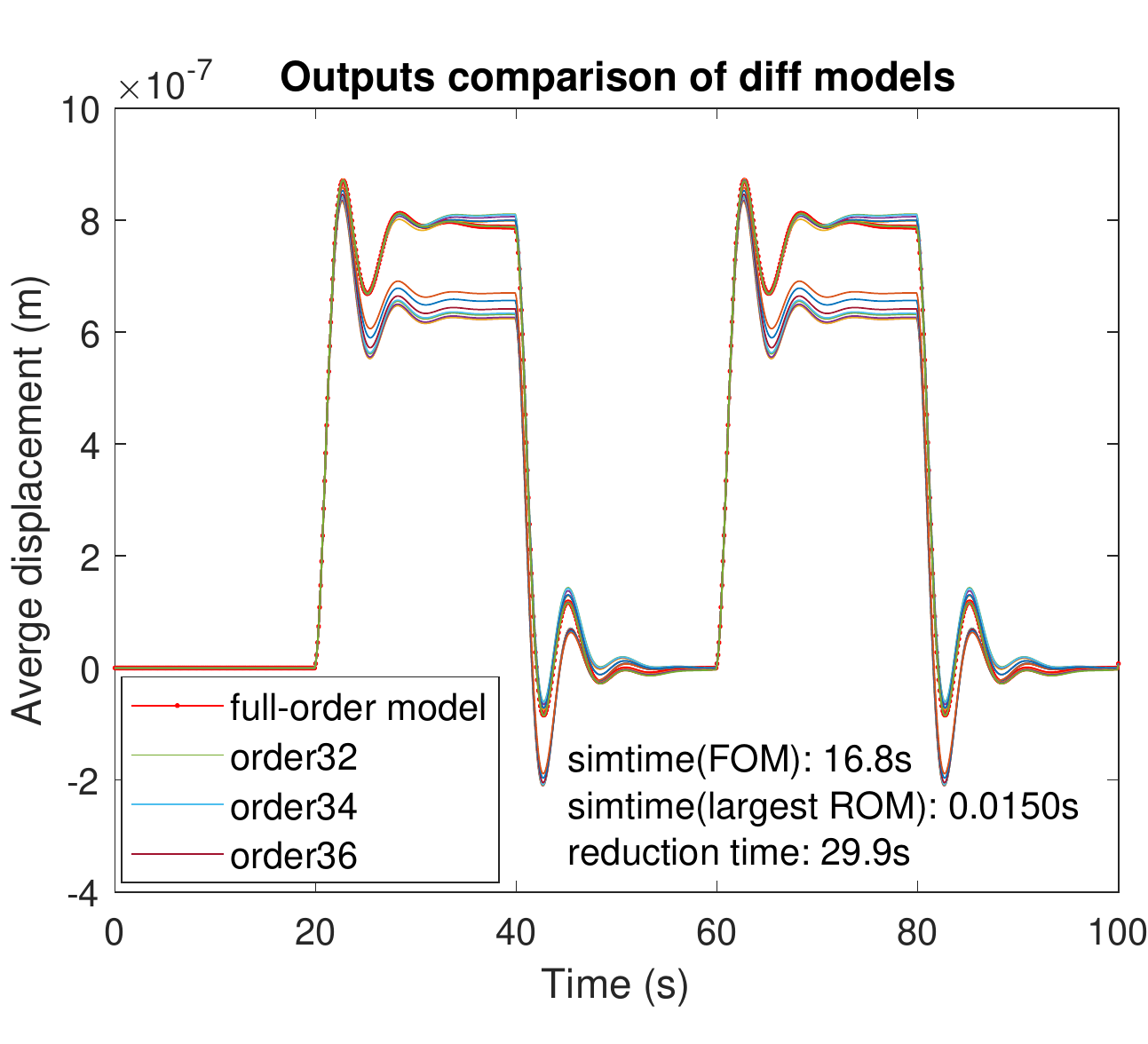}
		\subcaption{Case 1 \label{fig:behavior_bracket_case1}}		
	\end{minipage}  
	\begin{minipage}{0.46\textwidth}
		\centering
		\includegraphics[trim = 4cm 8cm 4.45cm 8cm,clip ,width=\linewidth]{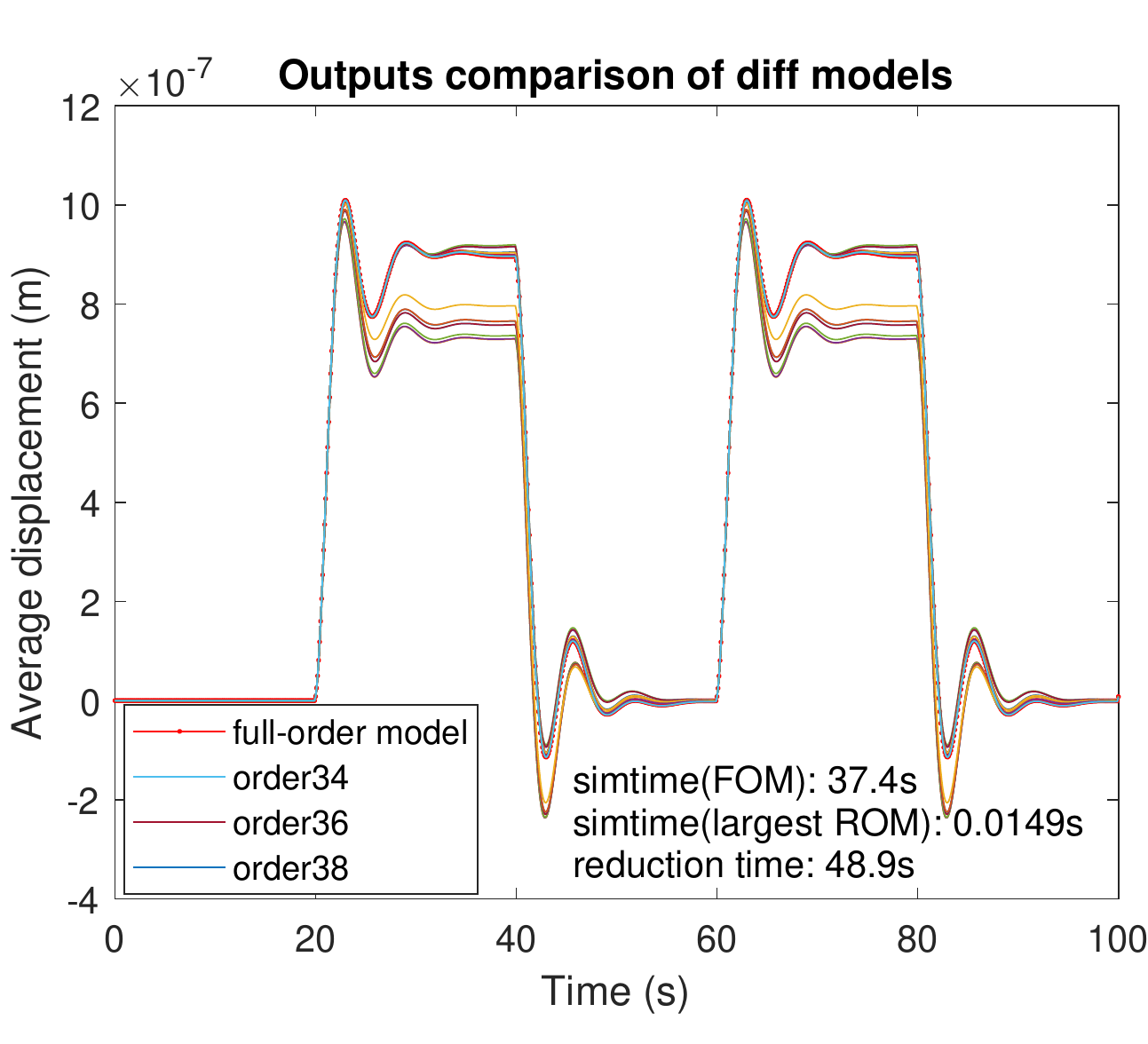}
		\subcaption{Case 2\label{fig:behavior_bracket_case2}}		
	\end{minipage}        
	\begin{minipage}{0.46\textwidth}
		\centering
		\includegraphics[trim = 4cm 8cm 4.45cm 8cm,clip ,width=\linewidth]{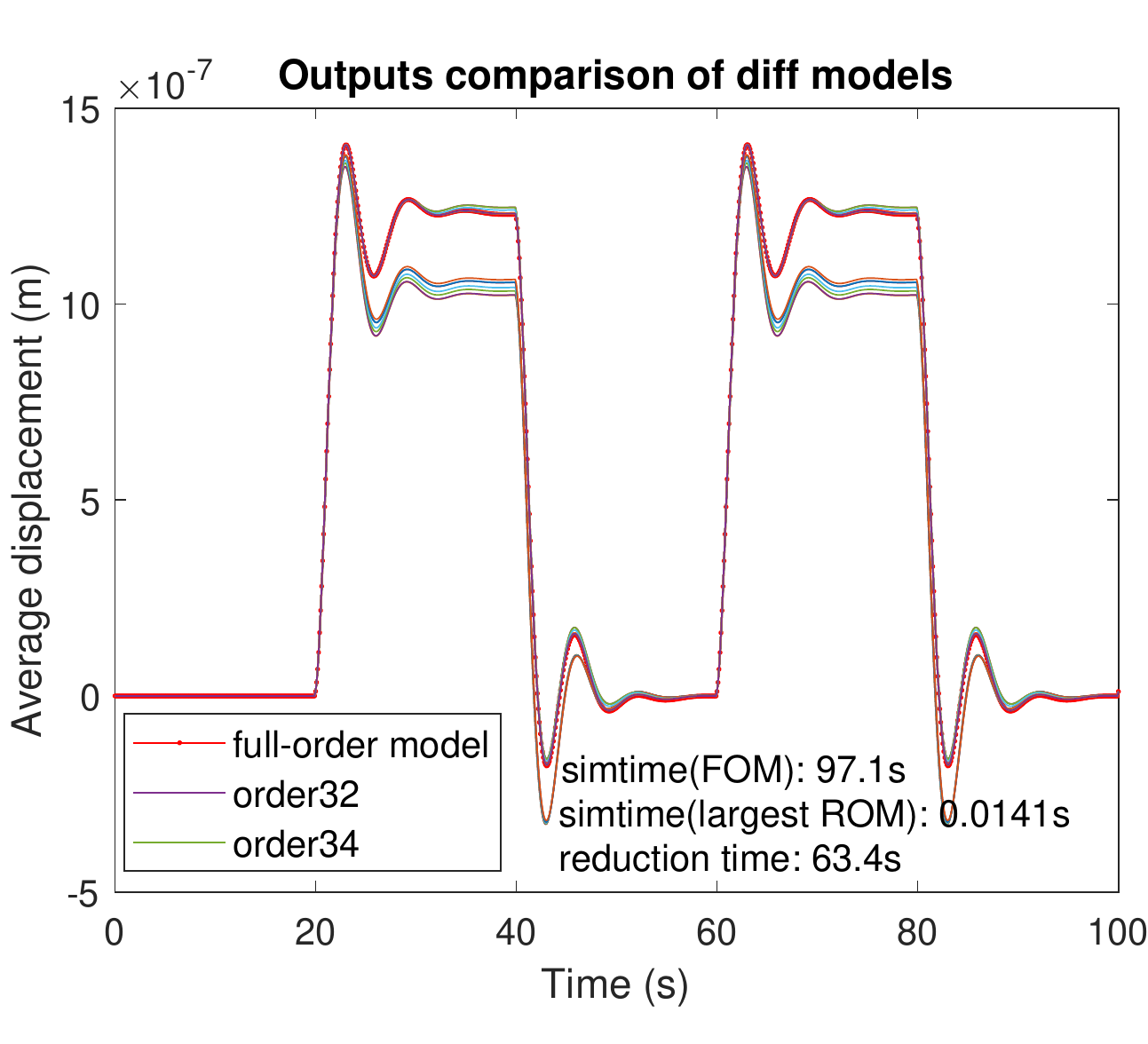}
		\subcaption{Case 3 \label{fig:behavior_bracket_case3}}		
	\end{minipage}        
	\begin{minipage}{0.46\textwidth}
		\centering
		\includegraphics[width=\linewidth]{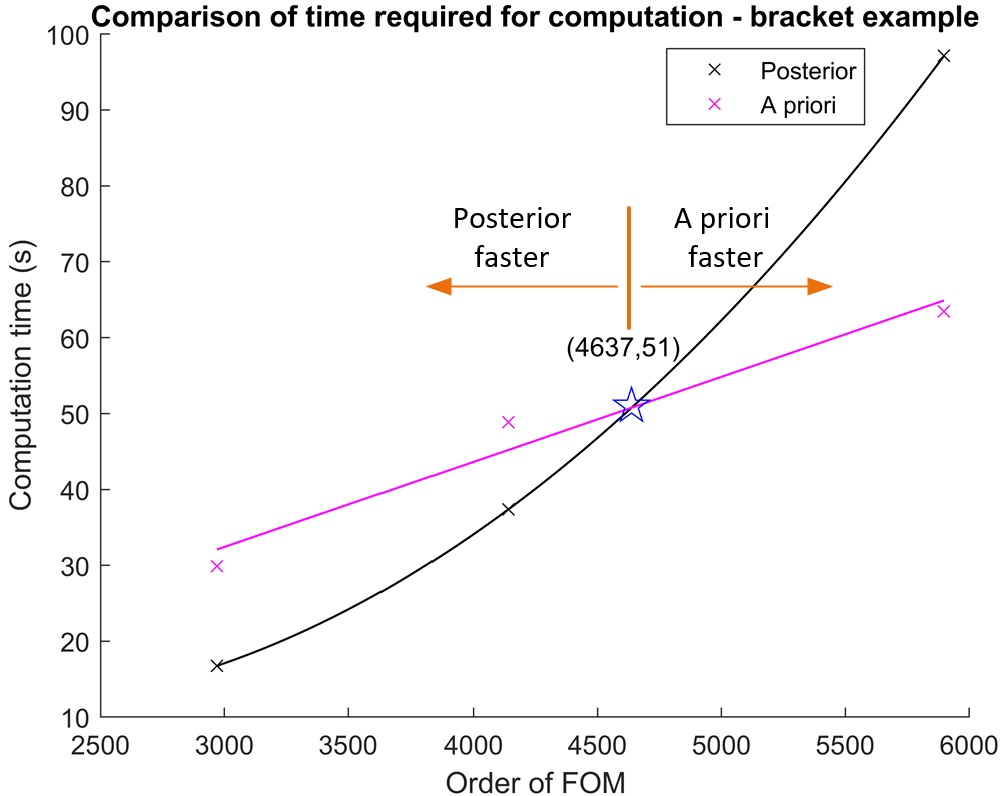}
		\subcaption{Time comparison \label{fig:bracket-time-comparison}}
	\end{minipage}                
	\caption{Comparison of the simulation results and computation time between DPM and ROMs - the bracket example \label{fig:behavior_bracket_cases}}	
\end{figure}

Through simulations of the DPM and all ROMs for three cases of the bracket, we have obtained two polynomial curves (Figure \ref{fig:bracket-time-comparison}) that fit the relation between computation time and the order of DPM for both a posterior solution comparison based on direct ODE simulations and our simulation-free scheme. The curves intersect at a certain point, indicating that if the order of the discretized DPM is lower than 4,637, comparing solutions through simulations is more time-efficient. However, if the order is higher than 4,637, our simulation-free scheme will be more time-efficient, with an accuracy loss of at most 1\%. It should be noted that if a smaller user-selected relative $\mathcal{H}_2$ error bound is used (i.e., <1\%), a new cross-over point that slightly shifts towards a larger order of DPM will be found.

\begin{figure}[!htb]
	\begin{minipage}{0.46\textwidth}
		\centering
		\includegraphics[trim = 4cm 8cm 4.5cm 8cm,clip ,width=\linewidth]{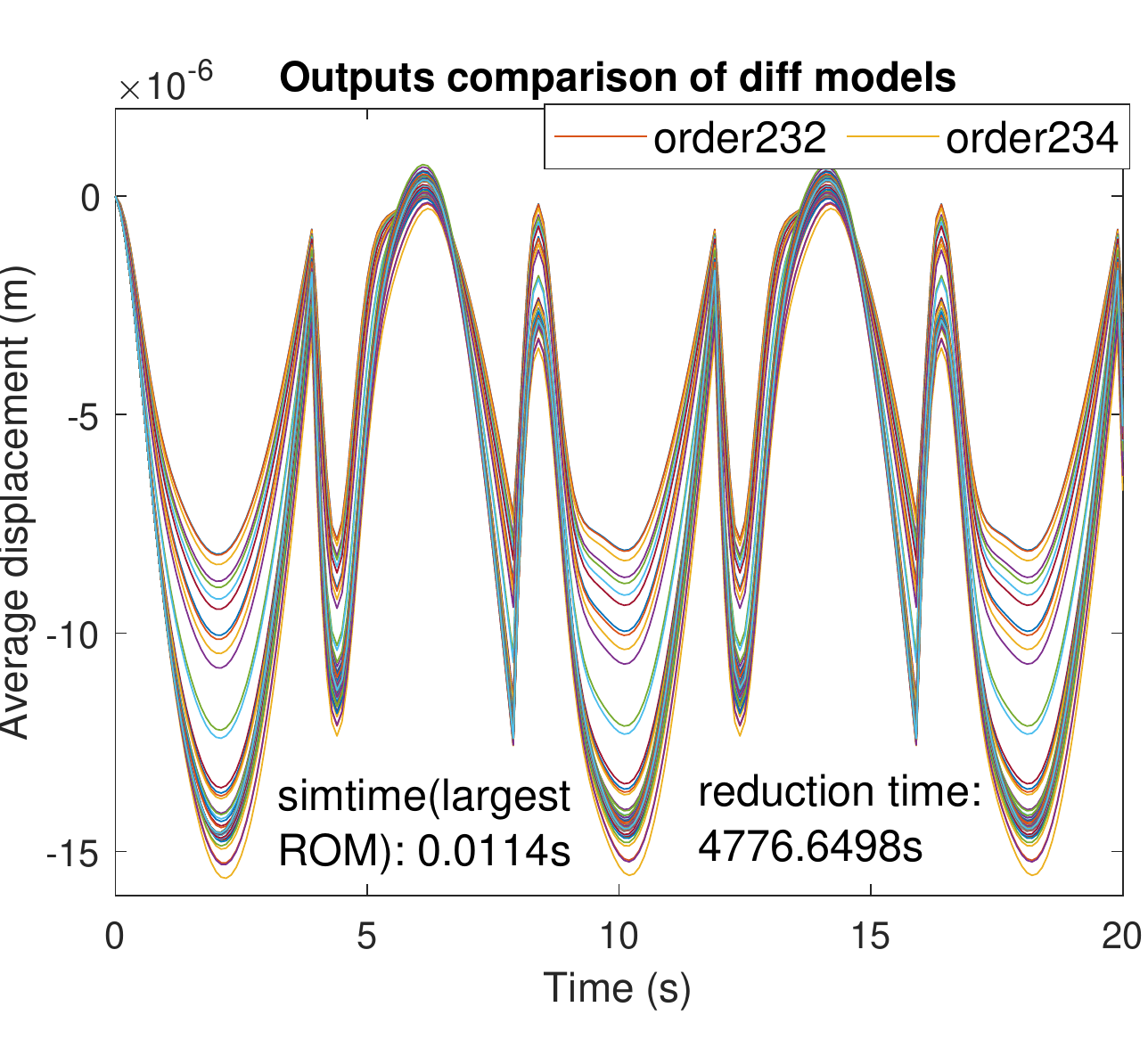}
		\subcaption{Solution comparison \label{fig:Behavior_bike_case1}}
	\end{minipage}  
	\begin{minipage}{0.46\textwidth}
		\centering
		\includegraphics[trim = 3.5cm 8cm 4.45cm 8cm,clip ,width=\linewidth]{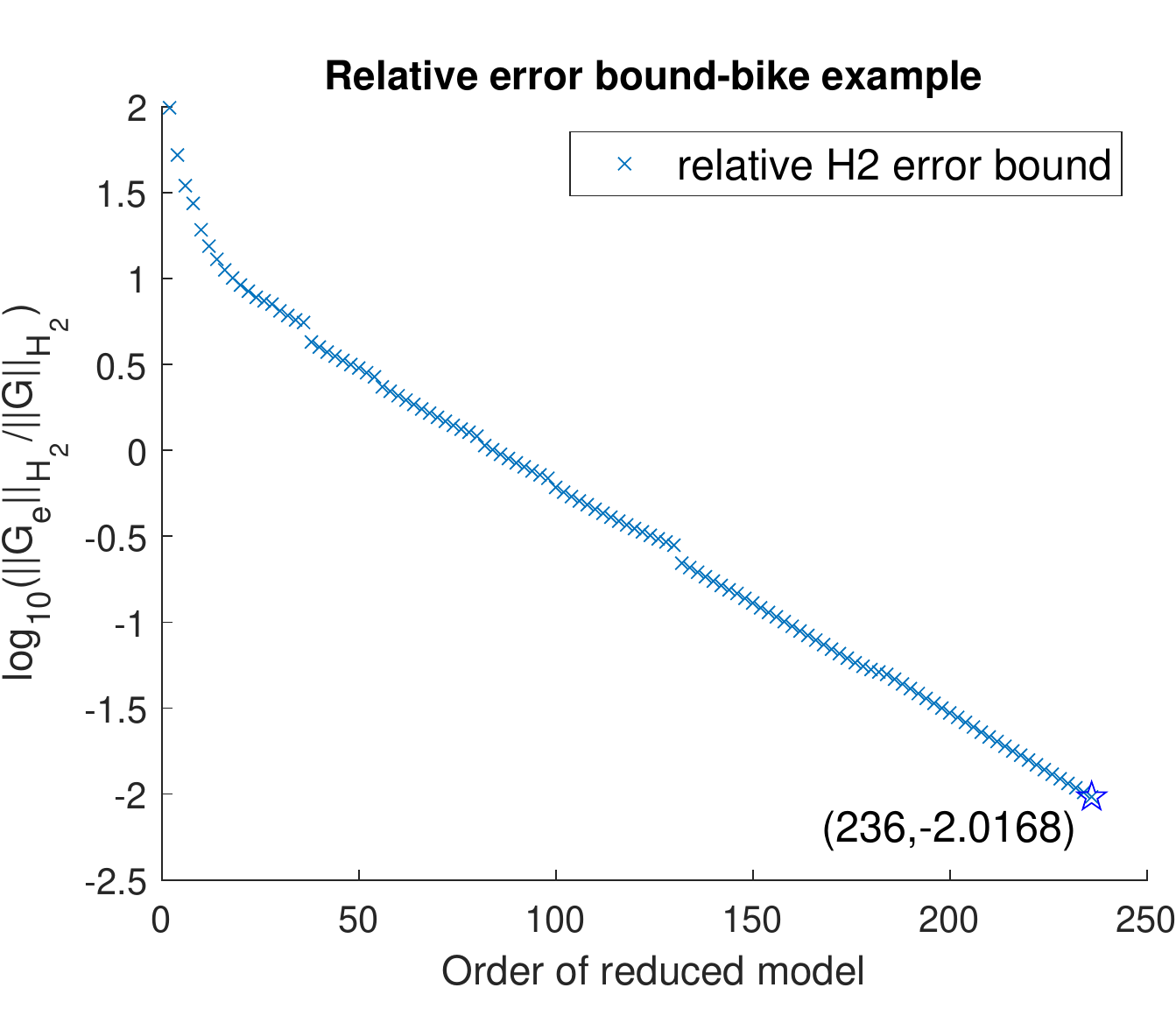}
		\subcaption{Relative ${{\mathcal{H}}_2}$ error bound\label{fig:error_bound_bike_case1}}		
	\end{minipage}             
	\caption{Comparison of the simulation results between DPM and ROM and a priori relative ${{\mathcal{H}}_2}$ error bound - frame example \label{fig:behavior_error_bound_bike_cases}}	
\end{figure}

\subsection{Surrogate DPM of the Frame and Error Analysis}
In a similar manner, we apply the SPARK+CURE approach to the frame design and compare the results of both DPM and ROMs (Figure \ref{fig:Behavior_bike_case1}). The computed relative $\mathcal{H}_2-$error bound is presented in Figure \ref{fig:error_bound_bike_case1}. The figure indicates that if the ROM order is greater than 236, the relative $\mathcal{H}_2$ error between the DPM and the surrogate DPM would be less than ${10^{ - 2.0168}} \approx 0.962\%$. Based on this, we can compute the a priori ${{\mathcal{H}}_2}$ relative error bound between the discretized DPM and the LPM as follows:

\begin{equation}
\begin{gathered}
  \frac{{{{\left\| {{{\mathbf{G}}_d}(s) - {{\mathbf{G}}_l}(s)} \right\|}_{{\mathcal{H}_2}}}}}{{{{\left\| {{{\mathbf{G}}_l}(s)} \right\|}_{{\mathcal{H}_2}}}}} \le {\text{ }}\frac{{{{\bar \varepsilon }_1} + {{\bar \varepsilon }_2}}}{{{{\left\| {{{\mathbf{G}}_l}(s)} \right\|}_{{\mathcal{H}_2}}}}}{\text{ }} ={\text{0.8229}} = {\bar \varepsilon }_{rel}  
\end{gathered} 
\end{equation}

\begin{figure}
	\begin{minipage}{0.46\textwidth}
		\centering
		\includegraphics[trim = 3.8cm 8cm 4.5cm 8cm,clip ,width=\linewidth]{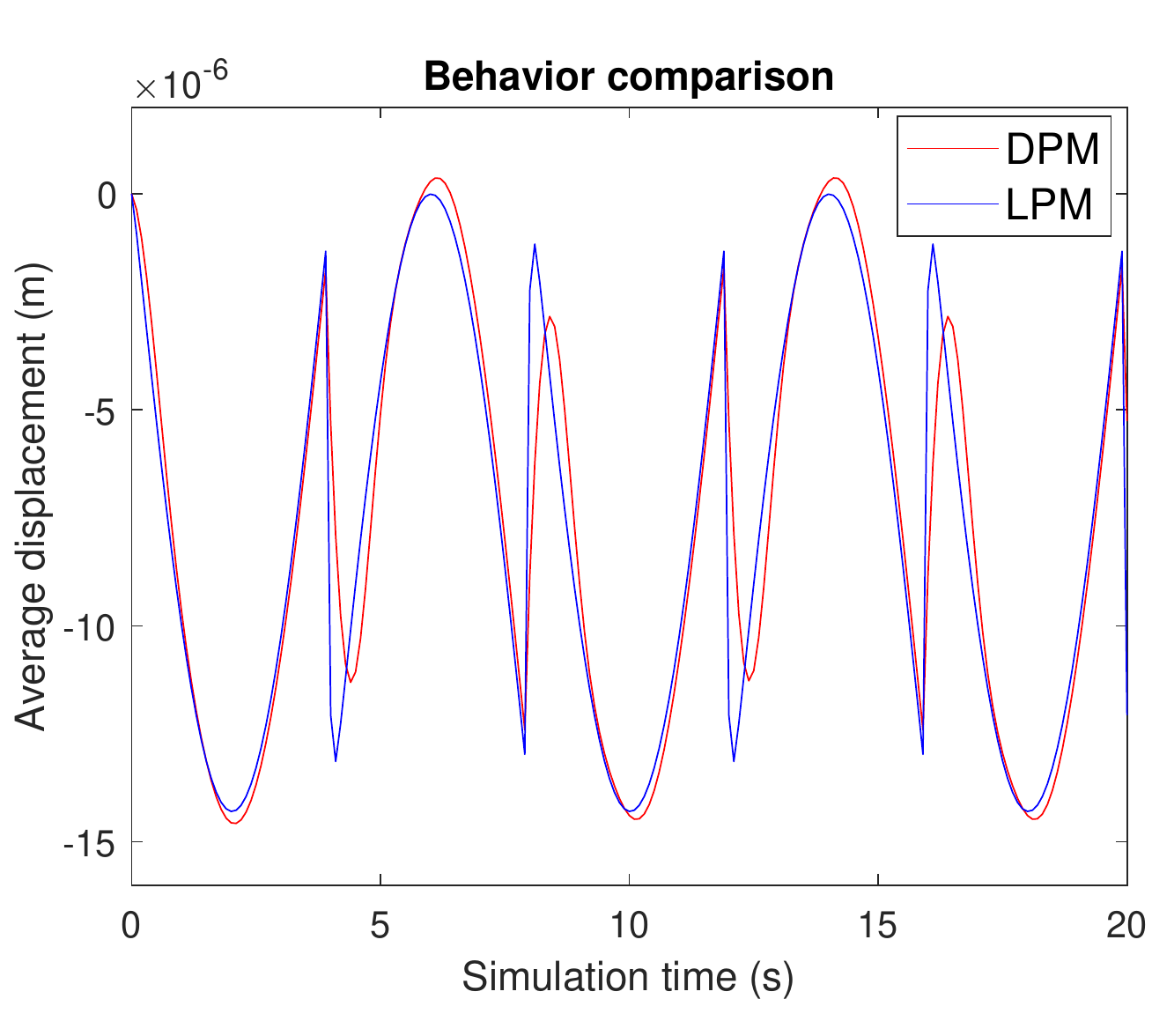}
		\subcaption{Solution comparison \label{fig:comparison_final_bike}}		
	\end{minipage}  
	\begin{minipage}{0.46\textwidth}
		\centering
		\includegraphics[trim = 3.6cm 8cm 4.5cm 8cm,clip ,width=\linewidth]{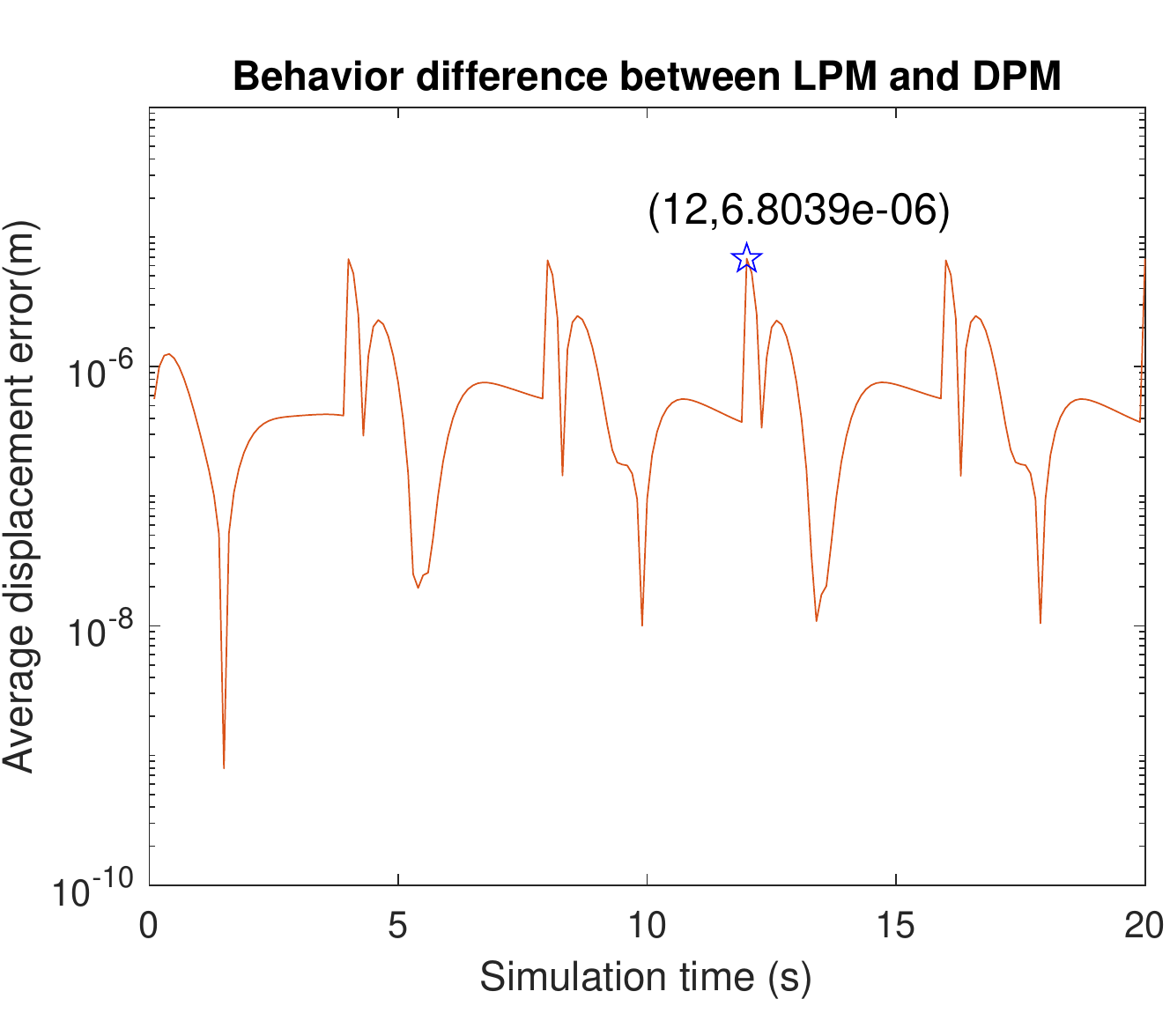}
		\subcaption{Error \label{fig:bound_final_bike}}		
	\end{minipage}                
	\caption{Numerical solution comparison between LPM and DPM - frame example\label{fig:results_bike}}	
\end{figure}

The agreement between the LPM and DPM can be determined by the user-defined threshold for the relative error. When the threshold is greater than $0.8229$, the models are in agreement. If the threshold is smaller than $0.8229$, the models are not in agreement. In Figure \ref{fig:comparison_final_bike}, we compare the simulation results of these two models, and their numerical solutions difference is depicted in Figure \ref{fig:bound_final_bike}. It can be observed that the difference is significantly smaller than the displacement of design models in most of the time steps.

\section{Conclusion}\label{sec:conclusion}

The paper introduces a model consistency concept and presents a simulation-free scheme for comparing the behavior of the LPM and the DPM. By leveraging the SPARK+CURE MOR method, the proposed approach can address the computational challenges associated with large-scale models. Our analysis of two mechanical models demonstrates that our simulation-free scheme provides relatively tight $\mathcal{H}_2$ a priori error bounds for behavior comparison. Furthermore, we show that our proposed scheme is more time-efficient than direct numerical simulation if the order of the discretized DPM is above a certain threshold, with increasing efficiency as the DPM order increases. The proposed approach can be applied to various engineering applications to enhance the system-based geometric design and facilitate integration between system-level and geometric designs.

However, we acknowledge that our framework has some limitations. One limitation is that the scheme to compare model solutions does not account for errors caused by spatial discretization of DPMs, and users must select an appropriate spatial discretization method and resolution. Additionally, the CURE scheme can only be applied to reduce the order of dissipative models, and the a priori guarantees provided by SPARK+CURE are only valid for LTI models. There are two ways to extend our proposed scheme to nonlinear models. The first approach is to divide the nonlinear MOR models into piecewise linear MOR models, and then apply the SPARK+CURE method, similar to the Trajectory PieceWise Linear (TPWL) Method \cite{rewienski2006model}. The second approach is to use MOR methods designed specifically for nonlinear models, such as the Symplectic MOR methods \cite{peng2016symplectic}, the MOR methods based on the proper orthogonal decomposition \cite{chaturantabut2010nonlinear}, and the Gaussian process regression  \cite{marzouk2009dimensionality}, etc. However, many of these methods do not offer rigorous a priori error guarantees.

\section*{Acknowledgments}
This material is based upon work supported by the Defense Advanced Research Projects Agency (DARPA) under Agreements No. HR00111990029 and HR00112090065. The responsibility for errors and omissions lies solely with the authors.

\bibliographystyle{asmeconf}  
\bibliography{asme2e}

\end{document}